\documentclass[sigconf]{acmart}

\usepackage{booktabs} 

\acmConference[CIKM'22]{The 45th International ACM SIGIR Conference on Research and Development in Information Retrieval}{October 17 - 22, 2022}{Georgia, USA}
\copyrightyear{2022}
\acmYear{2022}
\setcopyright{acmcopyright}\acmConference[CIKM '22]{Proceedings of the 31st ACM International Conference on Information and Knowledge Management}{October 17--21, 2022}{Atlanta, GA, USA}
\acmBooktitle{Proceedings of the 31st ACM International Conference on Information and Knowledge Management (CIKM '22), October 17--21, 2022, Atlanta, GA, USA}
\acmPrice{15.00}
\acmDOI{10.1145/3511808.3557405}
\acmISBN{978-1-4503-9236-5/22/10}

\usepackage{amsmath}
\usepackage{balance}
\usepackage{graphicx}
\usepackage{epsfig,epstopdf}
\usepackage{subfigure}
\usepackage{multirow}
\usepackage{booktabs}
\usepackage{color,xcolor}
\usepackage{url}
\usepackage{enumitem,balance,mathtools}
\usepackage{wrapfig}
\usepackage{euscript}
\usepackage{algorithm}
\usepackage{algorithmic}
\usepackage{ifpdf}
\usepackage{diagbox}
\usepackage{caption}
\usepackage{makecell}
\usepackage{subfigure}
\usepackage[leftcaption]{sidecap}
\usepackage{bm}
\usepackage{textcomp}
\usepackage{upgreek}
\usepackage{comment}

\usepackage{array}
\newcolumntype{N}{@{}m{0pt}@{}}

\usepackage{lipsum}

\newtheorem{definition}{Definition}

\newcommand{\minisection}[1]{\vspace{5pt}\noindent\textbf{#1.}}

\begin{document}
\title{Multi-Scale User Behavior Network for Entire Space\\ Multi-Task Learning}
\author{Jiarui Jin$^{1}$, Xianyu Chen$^{1}$, Weinan Zhang$^{1,*}$, Yuanbo Chen$^{2}$, Zaifan Jiang$^{2}$,\\ Zekun Zhu$^{2}$, Zhewen Su$^{2}$, Yong Yu$^{1}$.}
\affiliation{
\institution{$^1$Shanghai Jiao Tong University, $^2$Alibaba Group}
\country{China}
}
\email{{jinjiarui97, xianyujun, wnzhang, yyu}@sjtu.edu.cn, {yuanbo.cyb,zaifan.jzf, zekun.zhu, zhewen.su}@alibaba-inc.com}

	\renewcommand{\shortauthors}{J. Jin, et al.}
	\renewcommand{\shorttitle}{HEROES}

\ccsdesc[500]{Information systems~Recommender systems}

	\settopmatter{printacmref=false}
	
	\begin{abstract}
	Modelling the user's multiple behaviors is an essential part of modern e-commerce, whose widely adopted application is to jointly optimize click-through rate (CTR) and conversion rate (CVR) predictions.
	Most of existing methods are designed to encode the dependence between the behavior paths ``observation $\rightarrow$ click'' and ``click $\rightarrow$ conversion'' by combining the two-tower architecture with multi-task learning techniques to optimize multiple ranking objectives.
	However, such systems overlook the effect of two key characteristics of the user's behaviors:
	for each item list,
	(i) \emph{contextual dependence} refers to that the user's behaviors on any item are not purely determinated by the item itself but also are influenced by the user's previous behaviors (e.g., clicks, purchases) on other items in the same sequence;
	(ii) \emph{multiple time scales} means that users are likely to click frequently but purchase periodically.
	To this end, we develop a new multi-scale user behavior network named
	\textbf{\underline{H}}ierarchical r\textbf{\underline{E}}current \textbf{\underline{R}}anking \textbf{\underline{O}}n the \textbf{\underline{E}}ntire
	\textbf{\underline{S}}pace (\textsc{HEROES}) which incorporates the contextual information to estimate the user multiple behaviors in a multi-scale fashion. 
	Concretely, we introduce a hierarchical framework, where the lower layer models the user's engagement behaviors while the upper layer estimates the user's satisfaction behaviors.
	The proposed architecture can automatically learn a suitable time scale for each layer to capture the dynamic user's behavioral patterns.
	Besides the architecture, we also introduce the Hawkes process to form a novel recurrent unit which can not only encode the items' features in the context but also formulate the excitation or discouragement from the user's previous behaviors.
	We further show that HEROES 
	can be extended to build unbiased ranking systems through combinations with the survival analysis technique.
	Extensive experiments over three large-scale industrial datasets demonstrate the superiority of our model compared with the state-of-the-art methods.
	\end{abstract}

	\keywords{Multi-Scale User Modeling; Multi-Task Learning}

	\settopmatter{printacmref=false} 
	
	\maketitle
	
	{\fontsize{8pt}{8pt} \selectfont
	\textbf{ACM Reference Format:}\\
	Jiarui Jin, Xianyu Chen, Weinan Zhang, Yuanbo Chen, Zaifan Jiang, Zekun Zhu, Zhewen Su, Yong Yu. 2022. Multi-Scale User Behavior Network for Entire Space Multi-Task Learning. In \textit{Proceedings of the 31st ACM International Conference on Information and Knowledge Management, October 17--21, 2022, Atlanta, GA, USA}. ACM, New York, NY, USA, 10 pages.
	\url{https://doi.org/10.1145/3511808.3557405}}

	\section{Introduction} 
	\label{sec:intro}
\begin{figure}[t]
		\centering
		\includegraphics[width=1.0\linewidth]{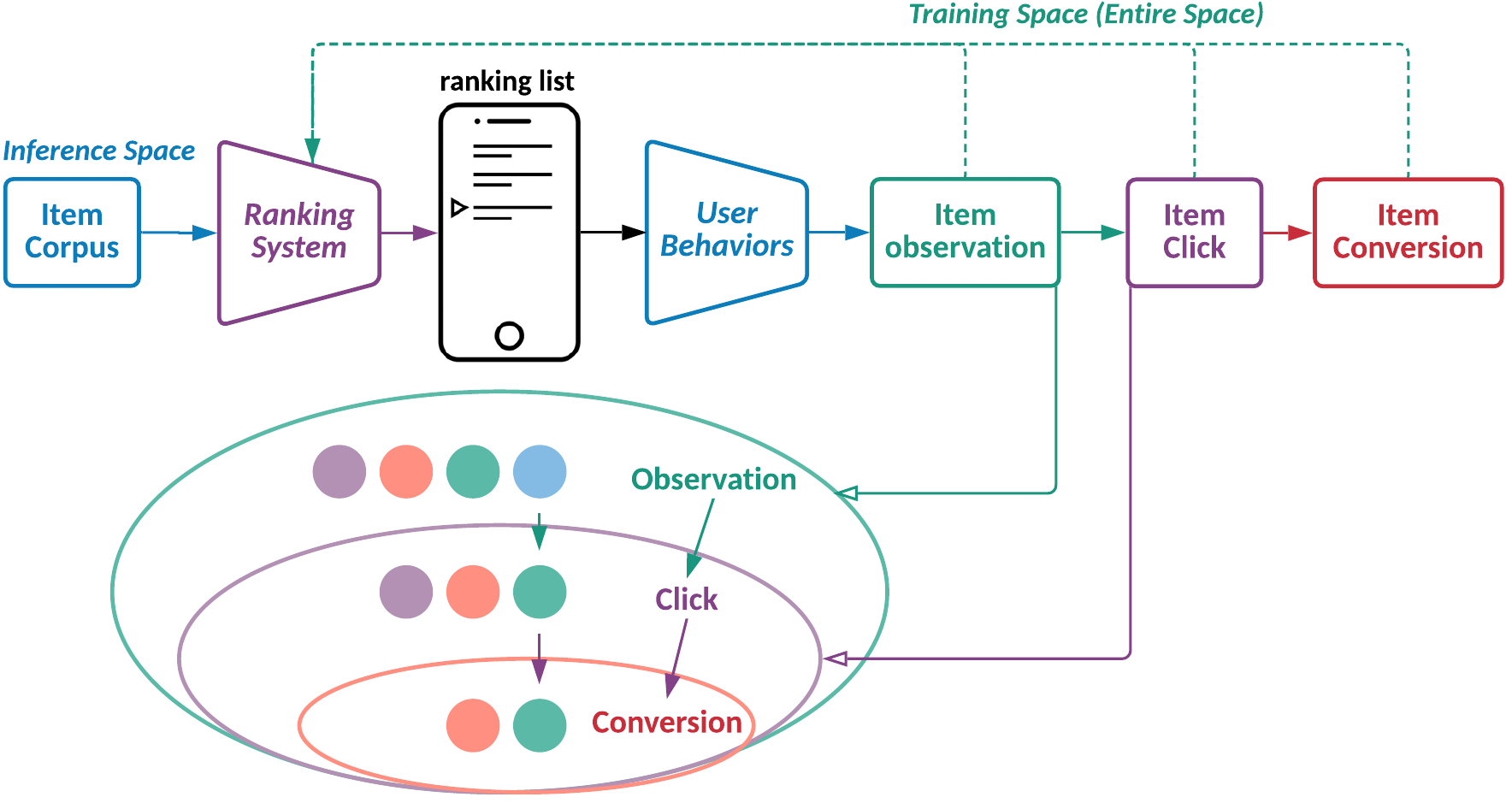}
		\vspace{-5.5mm}
		\caption{\small{Top diagram shows the pipeline information systems, which consists of a ranking system and the user multiple behaviors. Bottom figure illustrates the entire space behavior path ``observation $\rightarrow$ click $\rightarrow$ conversion''.}}
		\label{fig:bias}
		\vspace{-4.5mm}
	\end{figure}

	Implicit feedbacks from the user's behaviors are much important data sources for any personalized online service in information systems \citep{zhao2019recommending}.
	Such information systems often need to optimize multiple objectives at the same time \citep{ma2018modeling}.
	For example, when recommending videos to a user, the user is expected to not only click and watch the video but also to purchase it.
	Specifically, these behaviors can be categorized into two classes: (i) engagement behaviors, such as the user's observations and clicks; (ii) satisfaction behaviors, such as the user's watch time and purchases.
	As Figure~\ref{fig:bias} shows, a user would like to click and eventually purchase an observed item, which indicates a typical behavior path ``observation $\rightarrow$ click $\rightarrow$ conversion'' (i.e., ``engagement behaviors $\rightarrow$ satisfaction behaviors'') \citep{wen2019entire,ma2018entire}.
	We call it entire space behavior path.
    
    {
    \renewcommand{\thefootnote}{\fnsymbol{footnote}}
    \footnotetext[1]{Weinan Zhang is the corresponding author.}
    }

	Recent researches are mainly developed based on the existing multi-task learning techniques \citep{caruana1997multitask} to simultaneously learn multiple types of user behaviors, which can be roughly categorized into two directions.
	One direction \citep{zhao2015improving,ma2018entire,wen2019entire,meng2020incorporating} is to leverage the behavior decomposition which constructs the user's micro-actions as auxiliary information to promote the CTR and CVR predictions.
	Another line \citep{ma2018modeling,zhao2019recommending,wang2020m2grl} is to design an effective feature sharing strategy among the CTR and CVR prediction models.

	
	However, when modelling the user's multiple behaviors, almost all the existing papers have not well used, even may not be aware of, the following key characteristics of user behavior patterns:

\begin{itemize}[topsep = 3pt,leftmargin =10pt]
\item There exists \emph{contextual dependence} among the multiple behaviors in one list:
a user's behaviors on an item may affect her behaviors on its following items.
This influence would be either excitation or discouragement.
One example is that if a teen has purchased a hat, she then might not be interested in other hats.

\item Different behavior paths trigger with \emph{multiple time scales}.
For example, in Taobao e-commerce platform\footnote{\url{https://tianchi.aliyun.com/datalab/dataSet.html?dataId=408}}, the average time interval of clicks is 12.23, while that of purchases is 32.08, which indicates that the behavior paths ``observation $\rightarrow$ click'' and ``click $\rightarrow$ conversion'' happen with different time scales. 
\end{itemize}

\begin{figure}[t]
		\centering
		\includegraphics[width=1.0\linewidth]{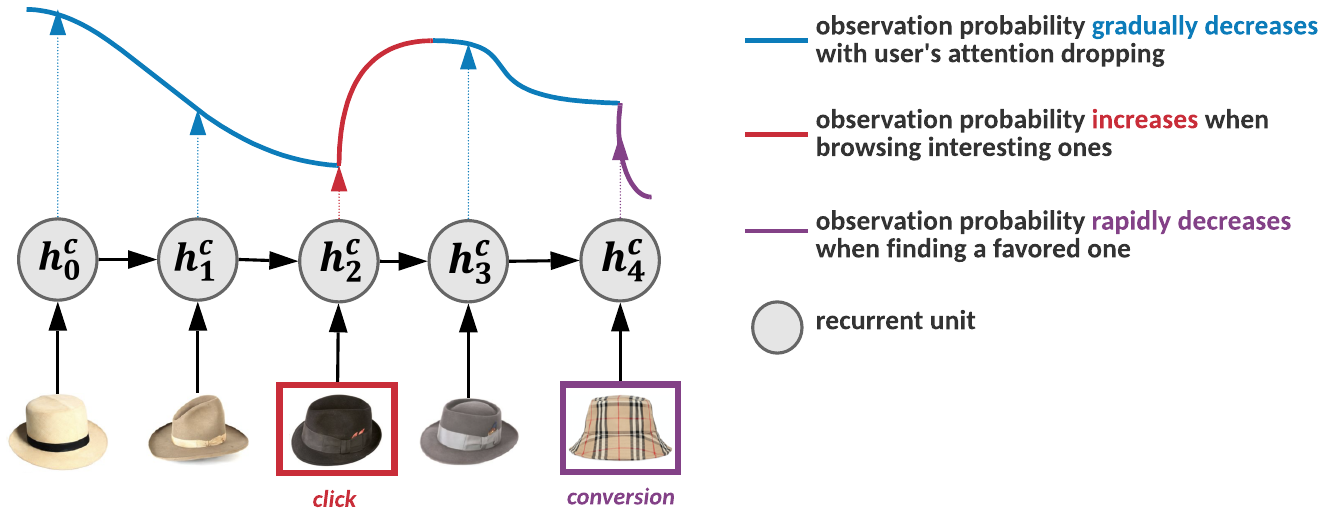}
		\vspace{-6mm}
		\caption{\small{An illustrated example of the characteristics of the user's behaviors:
		(i) \emph{contextual dependence}: 
		click occurs when browsing interesting items, which may encourage the observation, and conversion happens when finding a favored item, which may discourage the observation;
		(ii) \emph{multiple time scales}:
		conversions always happen after the user finding a favorite clicked item, and thus conversions occur less frequently than clicks.}
		}
		\label{fig:correlation}
		\vspace{-4mm}
	\end{figure}

\begin{figure*}[t]
		\centering
		\includegraphics[width=1.0\textwidth]{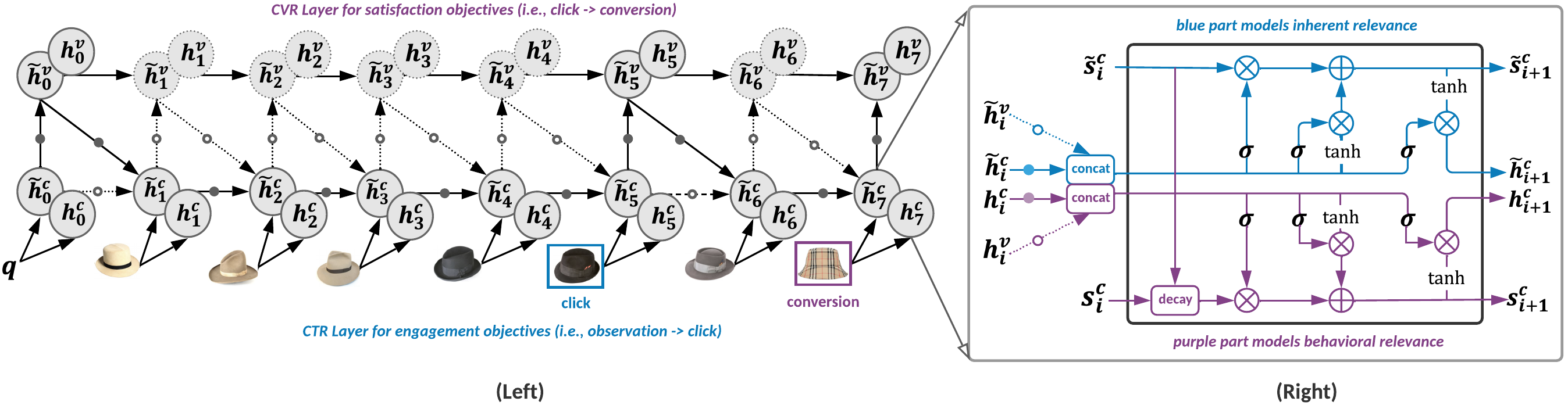}
		\vspace{-7mm}
		\caption{\small{An illustrated example of overall architecture (in the left part) and recurrent unit of HEROES (in the right part). 
		Left: in the lower layer (i.e., CTR layer) for engagement objective modeling, we model through the behavior path ``observation$\rightarrow$click''; while in the upper layer (i.e., CVR layer) for satisfaction objective modeling, we model through the behavior path ``click$\rightarrow$conversion''.
		Right: in the upper part, we incorporate the recurrent unit and the Hawkes process to form a HEROES unit, which can simultaneously model the inherent relevance and behavioral relevance by  mining the contextual item features and the user's previous behaviors.}
		}
		\label{fig:model}
		\vspace{-3mm}
	\end{figure*}

    An illustrated example of the above characteristics is provided in Figure~\ref{fig:correlation}.
    To this end, we propose a novel framework called 
    \textbf{\underline{H}}ierarchical r\textbf{\underline{E}}current \textbf{\underline{R}}anking \textbf{\underline{O}}n the \textbf{\underline{E}}ntire \textbf{\underline{S}}pace (\textsc{HEROES}) 
    to formulate the entire space user behavior path (i.e., ``observation $\rightarrow$ click $\rightarrow$ conversion'') in a multi-scale fashion.
    Concretely, as shown in Figure~\ref{fig:model}(Left), HEROES establishes a hierarchical architecture:
    the lower layer (i.e., CTR layer) estimates the engagement behaviors (i.e., ``observation $\rightarrow$ click''), while the upper layer (i.e., CVR layer) estimates the satisfaction behaviors (i.e., ``click $\rightarrow$ conversion'').
    We tweak the design of gates in \citep{chung2016hierarchical} and allocate them both within and between two layers, which allows HEREOS to automatically learn a suitable time scale for each layer.
    Besides the architecture, we then incorporate the neural Hawkes process \citep{mei2016neural} into the recurrent unit \citep{chung2016hierarchical} to form a new unit (called HEROES unit, as shown in Figure~\ref{fig:model}(Right)), which can not only encode the contextual item features but also model the excitation or discouragement brought from the user's past behaviors.
    By applying HEROES units into the hierarchical architecture, HEROES is able to simultaneously model all the top-down contextual contents (including the item features and the user's behaviors) and learn appropriate time scales (as Figure~\ref{fig:gate} depicts) for the entire space behavior path ``observation $\rightarrow$ click $\rightarrow$ conversion''.  
    
    
    Furthermore, we also show that HEROES can be seamlessly applied to the unbiased learning-to-rank task by incorporating with the survival analysis technique \citep{jin2020deep}.
    
	The major contributions of this paper can be outlined as follows.
    \begin{itemize}[topsep = 3pt,leftmargin =10pt]
    \item We propose a novel paradigm named HEROES, where we model the multiple user behaviors on entire space (i.e., ``observation $\rightarrow$ click $\rightarrow$ conversion'') in a multi-scale manner.
    \item We design a novel recurrent unit to take both the contextual items and the user's previous behaviors into consideration. 
    \item We show that our approach can be seamlessly used for unbiased ranking by incorporating with survival analysis technique. 
    \end{itemize}
	We conduct the comprehensive experiments on three industrial datasets, whose results exhibit that our method can learn an effective ranker over multiple objectives.
	To our knowledge, this work is the first work that simultaneously models the user's multiple behaviors in a multi-scale fashion.

	\section{Preliminary}
	\label{sec:back}
	\subsection{Learning-to-Rank with Multiple Objectives}
	The fundamental goal of learning-to-rank (LTR) scenarios is to learn a ranker $f$, which assigns a score to the item according to its feature.
	Then, the item list concerning a query $q$ is provided in the descending order of their scores.
    Let $\mathcal{D}_q$ denote the set of items associated with $q$, $d_i$ denote the $i$-th item in $\mathcal{D}_q$ and $\bm{x}_i$ denote the feature vector of $d_i$.
	Let $b_i$ represent the score of $d_i$.
	For simplicity, we only consider the binary score here, i.e., $b_i=0$ or $b_i=1$.
	One can easily extend it to the multi-level scores.
	We can describe the risk function as
	\begin{equation}
	\label{eqn:base}
	    \mathcal{R}(f) = \sum_q \sum_{d_i\in \mathcal{D}_q} L(f(\bm{x}_i),b_i),
	\end{equation}
	where $f$ denotes a ranker, and $L(f(\bm{x}_i),b_i)$ denotes a loss function. 
	The goal of LTR is to find the optimal ranker $f^*$ that minimizes the loss function.
	In the CTR prediction, the ranker is learned with implicit feedbacks containing click labeled data (i.e., the score $b_i$ of $d_i$ in Eq.~(\ref{eqn:base}) is replaced by its click signal $c_i$); while in the CVR prediction, $b_i$ of $d_i$ in Eq.~(\ref{eqn:base}) is replaced by its conversion signal $v_i$.
	Here $v_i$ is a binary value that denotes whether the conversion event occurs at $d_i$.
	
	The goal in this paper is to jointly optimize the CTR and CVR predictions.
    Formally, the input is $\mathcal{D}_q$ organized as a set of data samples $\{(\bm{x},c,v,I)\}$, where $\bm{x}$ is the item features, $I$ is the length of the item list, $c$ is the click signal, and $v$ is the conversion signal.
	The output is the predictions of click and conversion probabilities of each item $d_i$ (i.e., $P(c_i=1|\bm{x}_i)$ and $P(v_i=1|\bm{x}_i)$).
	
	\subsection{Relevance for Behavior Dependence}
	Note that the user's multiple behaviors naturally have dependence among each other (a.k.a., behavior path ``observation $\rightarrow$ click $\rightarrow$ conversion'').
	Hence, instead of separately modelling the click and conversion, the current prevailing approaches introduce the  
    post-click relevance (denoted as $r^v_i$ for $d_i$) (so called the post-view conversion rate \citep{ma2018entire,wen2019entire}), which is defined as
	\begin{equation}
	\label{eqn:post}
	P(r^v_i=1|\bm{x}_i) \coloneqq P(v_i=1|c_i=1;\bm{x}_i)=\frac{P(v_i=1|\bm{x}_i)}{P(c_i=1|\bm{x}_i)},
	\end{equation}
	which allows the model to first separately estimate $P(c_i=1|\bm{x}_i)$ and $P(r^v_i=1|\bm{x}_i)$ and then recover $P(v_i=1|\bm{x}_i)$ by computing the production.
	
	As the post-click relevance (i.e., $r^v_i$ for $d_i$) is built based on the behavior path ``click $\rightarrow$ conversion'', we can similarly define the prior-click relevance (denoted as $r^c_i$ for $d_i$)
	as
	\begin{equation}
	\label{eqn:prior}
	    P(r^c_i=1|\bm{x}_i) \coloneqq P(c_i = 1 | o_i =1; \bm{x}_i) = \frac{P(c_i = 1 | \bm{x}_i)}{P(o_i = 1 | \bm{x}_i)},
	\end{equation}
	which reflects a simple fact that a user clicks ($c_i=1$) the item $d_i$ only when it is both observed ($o_i=1$) and perceived as prior-click relevant ($r^c_i=1$).
	Eq.~(\ref{eqn:prior}) has been widely used in unbiased LTR researches \citep{joachims2017unbiased,wang2016learning,wang2018position,ai2018unbiaseda,jin2020deep}, whose goal is to discover the prior-click relevance from the biased click data.
	In contrast, biased LTR refers to directly regarding the click as the prior-click relevance, where the prior-click relevance is defined as
	\begin{equation}
	\label{eqn:biasprior}
	    P(r^c_i= 1|\bm{x}_i)\coloneqq P(c_i=1|\bm{x}_i).
	\end{equation}

	\section{HEROES}
	\label{sec:model}
	In this section, we present the HEROES in the context of the biased LTR (i.e., using Eq.~(\ref{eqn:biasprior})) where we first describe our architecture design and then introduce the proposed HEROES unit to construct the architecture.
	Finally, we show the loss function for biased LTR.
	
	\subsection{HEROES Architecture}
	\label{sec:architecture}
	In HEROES, we first divide the entire space behavior path into two layers: ``observation $\rightarrow$ click'' in the CTR layer and ``click $\rightarrow$ conversion'' in the CVR layer, and then model the user's multiple behaviors both within and between two layers.
	
	We begin by introducing the definition of the inherent relevance:
	\begin{definition}
	\label{def:inherent}
	\emph{\textsc{(Inherent Relevance)}}
	For each item $d_i$, we define a binary variable $\widetilde{r}_i$ as the inherent relevance.
	Specifically, $\widetilde{r}_i^c$ and $\widetilde{r}_i^v$ are the inherent relevance to motivate a user to click and purchase $d_i$ respectively, both of which are solely determinated by the item features and are free of the effect from all the external factors such as the contextual items and the user's past behaviors.
	\end{definition}
	
	In contrast, we call $r^v_i$ in Eq.~(\ref{eqn:post}) and $r^c_i$ in Eq.~(\ref{eqn:prior}) \emph{behavioral relevance} for conversion and click of $d_i$ respectively, which are affected by the external factors.
	To estimate $\widetilde{r}_i$ and $r_i$\footnote{For simplicity, we use  $r_i$ to denote both $r^c_i$ and $r^v_i$; and similar notations for $\widetilde{r}_i$, $h_i$, $f_\theta$, $z$.} for each item $d_i$, we further introduce $\widetilde{h}_i$ and $h_i$ which are defined as
	\begin{equation}
	\begin{aligned}
	\widetilde{h}^c_i &\coloneqq P(\widetilde{r}^c_i=1), \
	\widetilde{h}^v_i \coloneqq P(\widetilde{r}^v_i=1);\\
	h^c_i &\coloneqq P(r^c_i=1), \
	h^v_i \coloneqq P(r^v_i=1).
	\end{aligned}
	\end{equation}

	\begin{table}[t]
    \caption{\small{A summary of notations regarding the item $d_i$}. 
    }  
    \label{tab:notation}
    \vspace{-4mm}
    \begin{center}  
    \begin{tabular}{p{2cm}<{\centering}p{6cm}<{\centering}}
    \toprule
    \textbf{Notations} & \textbf{Explanations} \\
    \midrule
    $c_i, v_i$ & Click, conversion (implicit feedback)\\
    \midrule
    $r^c_i$ & Prior-click (behavioral) relevance (see Eq.~(\ref{eqn:prior}) for unbiased LTR, and Eq.~(\ref{eqn:biasprior}) for biased LTR) \\
    \midrule
    $r^v_i$ & Post-click (behavioral) relevance (see Eq.~(\ref{eqn:post})) \\
    \midrule
    $\widetilde{r}^c_i,\widetilde{r}^v_i$ & Inherent relevance (see \textsc{definition}~\ref{def:inherent})\\
    \midrule
    $h^c_i,h^v_i,\widetilde{h}^c_i,\widetilde{h}^v_i$ & Probability of $r^c_i=1,r^v_i=1,\widetilde{r}^c_i=1,\widetilde{r}^v_i=1$\\
    \bottomrule
    \end{tabular}  
    \end{center} 
    \vspace{-3mm}
\end{table}

	\minisection{Intra-Layer Behavior Modelling}
	In each layer (i.e., behavior path), since the user's behaviors on each item $d_i$ (i.e., $h_i$) can be either excited or discouraged by the user's previous behaviors, we apply the
	Hawkes process \citep{hawkes1971spectra,embrechts2011multivariate} to formulate the behavioral relevance as
	\begin{equation}
	\label{eqn:hawke}
	h_i := \widetilde{h}_i + \sum_{j\leq i}\lambda_j\exp{(-\delta_j(t_i-t_j))},
	\end{equation}
	where $\lambda_j \in\mathbb{R}$ is the \emph{learnable} degree to which the user's behaviors (e.g., click or conversion) on item $d_j$ initially excite (when $\lambda_j>0$) or discourage (when $\lambda_j<0$) that on item $d_i$; and $\delta_j > 0$ is the \emph{learnable} decay rate of the excitation or discouragement. 
	In other words, when studying $d_i$, as the time interval $t_i-t_j$ increases, its behavioral relevance $h_i$ might both rise and fall (conditioned on the effects from the intermediate items, i.e., $\{\lambda_j\}_{j\leq i}$), but eventually approach its inherent relevance $\widetilde{h}_i$, as the influences from previous behaviors on $d_j$ will decay toward 0 at rate $\delta_j > 0$.
	Here, $t_i$, $t_j$ are the behavior occurrence time in continuous space and can be roughly approximated by $i$ and $j$ in discrete time space.

	Note that Eq.~(\ref{eqn:hawke}) can be regarded as a conceptual formulation, as it can not guarantee $h_i, \widetilde{h}_i\in[0, 1]$.
	We will later introduce our HEROES unit design which implements Eq.~(\ref{eqn:hawke}) to the contextual item feature modelling in Section~\ref{subsec:unit}.
	\minisection{Inter-Layer Behavior Modelling}
	As introduced above, for each item $d_i$, there are two behavioral factors: the prior-click relevance $h^c_i$ representing how likely the user would click $d_i$ after observing it in the CTR layer, and the post-click relevance $h^v_i$ representing how likely the user would purchase $d_i$ after clicking it in the CVR layer.
	We explicitly model the correlations across these layers as followings.
	\begin{equation}
	\label{eqn:relation}
	h^v_i = \phi(h^c_i),\ h^c_{i+1} = \psi(h^v_i),
	\end{equation}
	where $\phi(\cdot)$ and $\psi(\cdot)$ denote parameterized mapping functions, 
	and we will specify them in later Section~\ref{subsec:unit}.
	The intuition behind Eq.~(\ref{eqn:relation}) is straightforward:
	The former equation represents the case where a user may click on an item because its abstract content (e.g., title) is interesting (i.e., $h^c_i$) and purchase it after carefully checking whether its detailed information (e.g., description) is relevant (i.e., $h^v_i$). 
	The latter equation shows the case where a user finds and purchases a favored item (i.e., $h^v_i$), and then it is likely for the user to stop browsing and not click its following items (i.e, $h^c_{i+1}$).
	
	\begin{figure}[t]
		\centering
		\includegraphics[width=1.0\linewidth]{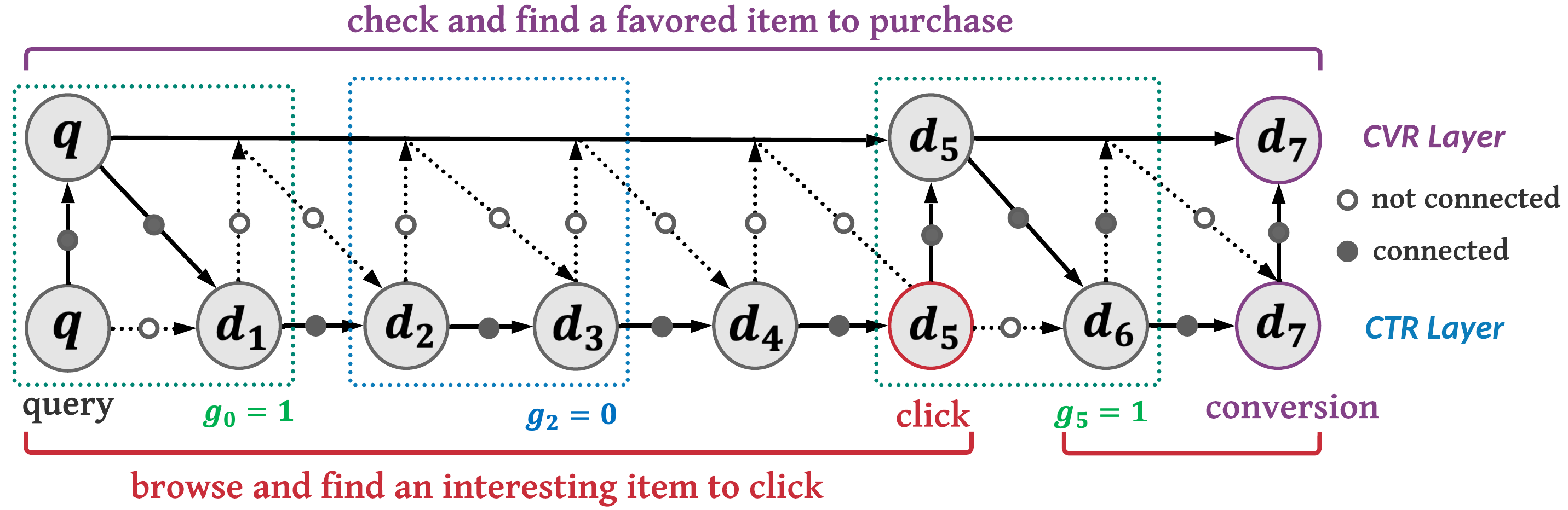}
		\vspace{-6mm}
		\caption{\small{An illustrated example of the gate mechanism: $g_i=0$ denotes no click at $d_i$ where the contextual information is passed via ``observation$\rightarrow$click'' in the CTR layer; and $g_i=1$ denotes a click at $d_i$ where the contextual information is summarized in the CTR layer and is further passed via ``click$\rightarrow$conversion'' in the CVR layer.}}
		\label{fig:gate}
		\vspace{-3mm}
	\end{figure}
	
	According to Eq.~(\ref{eqn:relation}), we introduce a hierarchical architecture, whose formulation can be written as
	\begin{equation}
	\label{eqn:rnn}
	\begin{aligned}
	h^c_i, \widetilde{h}^c_i & = f^c_\theta (h^c_{i-1}, h^v_{i-1}, \widetilde{h}^c_{i-1}, \widetilde{h}^v_{i-1}; \bm{x}_q);\\
	h^v_i, \widetilde{h}^v_i & = f^v_\theta (h^c_{i}, h^v_{i-1}, \widetilde{h}^c_{i}, \widetilde{h}^v_{i-1};\bm{x}_q),
	\end{aligned}
	\end{equation}
	where $f^c_\theta$ and $f^v_\theta$ are the recurrent functions in the CTR and CVR layers that take the contextual information $\bm{x}_q$\footnote{For each recurrent unit for $i$-th document $d_i$, we concatenate the one-hot embedding vector of its position (i.e., $i$) and its document feature (i.e., $\bm{x}_i$) as the input. 
	Considering that recurrent network will encode all the contextual information in query $q$ (i.e., $D_q$), we use $\bm{x}_q$ for simplicity.} as the input, and output $h^c_i$, $\widetilde{h}^c_i$ and $h^v_i$, $\widetilde{h}^v_i$.
    
    \minisection{Gate Mechanism}
    Besides the aforementioned intra- and inter-layer behavior modelling which are proposed to incorporate the contextual information, we further introduce a gate mechanism to explicitly discover the hidden structure.
    Specifically, the CTR layer is expected to model the contextual information in each segment following the behavior path ``observation$\rightarrow$click'' (i.e., starting at a user observing an item and ending at the user clicking an item).
    Similarly, the CVR layer is expected to model the contextual information in each segment following the behavior path ``click$\rightarrow$conversion'' (i.e., starting at a user clicking an item and ending at the user purchasing an item).

    To this end, we define $g_i$ as a boundary detector as
    \begin{equation}
	\label{eqn:bound}
	g_i = \left\{
	\begin{aligned}
	1,  \ & \ \text{if } P(c_i=1) > 0.5\\
	0. \ & \ \text{otherwise}
	\end{aligned}
	\right.,
	\end{equation}
	We illustrate how $g_i$ influences the hierarchical structure in Figure~\ref{fig:gate}:
	If there is no click at $d_i$ (i.e., $g_i=0$), then the contextual information of $d_i$ should pass through the behavior path ``observation$\rightarrow$click'' in the CTR layer; otherwise (i.e., $g_i=1$), the contextual information of $d_i$ should be summarized in the CTR layer and pass through the behavior path ``click$\rightarrow$convsersion'' in the CVR layer.


	\subsection{HEROES Unit}
	\label{subsec:unit}
	For each item $d_i$, HEROES unit models model inherent relevance $\widetilde{h}_i$ and behavioral relevance $h_i$ in a recurrent fashion which incorporate the contexts both within (i.e., intra-layer) and across (i.e., inter-layer) the CTR and CVR layers.
	
    \minisection{Intra-Layer Mechanism}
    Let $\widetilde{s}_i$ and $s_i$ denote the unit states in each cell which correspond to $\widetilde{h}_i$ and $h_i$.
	To achieve the gate mechanism in Figure~\ref{fig:gate}, we first recursively compute the states in the CTR layer as:
	\begin{equation}
	\label{eqn:cellctr}
	\begin{aligned}
	\widetilde{s}^c_i &= \left\{
	\begin{aligned}
	\widetilde{\text{f}}^c_i \odot \widetilde{s}^c_{i-1} + \widetilde{\text{i}}^c_i \odot \widetilde{\text{g}}^c_i, \ \langle 1 \rangle \\
	\widetilde{\text{i}}^c_i \odot \widetilde{\text{g}}^c_i, \ \langle 2 \rangle
	\end{aligned}
	\right. 
	;\ 
	s^c_i = \left\{
	\begin{aligned}
	\text{f}^c_i \odot s^c(t_{i-1}) + \text{i}^c_i \odot \text{g}^c_i, \ \langle 1 \rangle \\
	\text{i}^c_i \odot \text{g}^c_i, \ \langle 2 \rangle
	\end{aligned}
	\right.
	\end{aligned},
	\end{equation}
	where $\odot$ denotes element-wise product operation.
	$\langle 1 \rangle\langle 2\rangle$ respectively denote the conditions: $g_{i-1}=0, g_{i-1}=1$.
	Here, $\text{f}_i$ (i.e., $\text{f}_i^c, \text{f}_i^v$, $\widetilde{\text{f}}_i^c$, $\widetilde{\text{f}}_i^v$); $\text{i}_i$ (i.e., $\text{i}_i^c, \text{i}_i^v$, $\widetilde{\text{i}}^c$, $\widetilde{\text{i}}^v$); $\text{o}_i$ (i.e., $\text{o}_i^c, \text{o}_i^v$ $\widetilde{\text{o}}_i^c, \widetilde{\text{o}}_i^v$) are forget, input, output gates, and $\text{g}_i$ (i.e., $\text{g}_i^c, \text{g}_i^v$ $\widetilde{\text{g}}_i^c, \widetilde{\text{g}}_i^v$) is a cell proposal vector, whose calculations will be later introduced in Eq.~(\ref{eqn:cellgate}).
	
	We call the operation under $\langle 1\rangle$ \emph{update}, and it under $\langle 2\rangle$ \emph{summarize}.
	When there is no click at $d_{i-1}$, then there should the contextual information from $d_{i-1}$ passing to $d_{i}$, and thus \emph{update} operation will update the states at $d_i$.
	Otherwise (i.e., there is a click at $d_{i-1}$), then \emph{summarize} operation will summarize the contextual information in the current segment to update it in the CVR layer (see Eq.~(\ref{eqn:cellcvr})) and reinitialize the state for the next segment. 
	
	In \emph{update} operation, we introduce $s^c(t)$ to formulate the Hawkes process in Eq.~(\ref{eqn:hawke}) as 
	\begin{equation}
	\label{eqn:st}
	s^c(t)=\widetilde{s}^c_{i+1}+(s^c_{i+1}-\widetilde{s}^c_{i+1})\exp{(-\delta_{i+1}(t-t_i))} \text{ for } t\in (t_i,t_{i+1}].
	\end{equation}
	Here, $\delta_{i+1}=f_\gamma(\text{MLP}(y_{i+1}||h^c_{i+1}))$ where $y_{i+1}$ is a multi-hot embedding vector representing the user's behavior on item $d_{i+1}$, and $||$ denotes the concatenation operation. 
	If $y_{i+1}$ is not available in some cases, we can directly apply $\delta_{i+1}=f_\gamma(\text{MLP}(h^c_{i+1}))$ instead.
	Here, we follow \citep{mei2016neural} to define $f_\gamma(x)$ as $f_\gamma(x)\coloneqq\gamma \log(1+\exp{(x/\gamma)})$ where $\gamma$ is set as 5 in our experiment.
	
	Then, the hidden states $\widetilde{h}^c_i$ and $h^c_i$ can calculated by 
	\begin{equation}
	\label{eqn:hiddenctr}
	\begin{aligned}
	\widetilde{h}^c_i = \widetilde{\text{o}}^c_i \odot \text{tanh}(\widetilde{s}^c_i), \ 
	h^c_i = \text{o}^c_i \odot \text{tanh}(s^c_i)
	\end{aligned}
	.
	\end{equation}
	
	Similar as Eq.~(\ref{eqn:cellctr}), the states in the CVR layer are recursively updated following
	\begin{equation}
	\label{eqn:cellcvr}
	\begin{aligned}
    \widetilde{s}^v_i &= \left\{
	\begin{aligned}
	\widetilde{\text{f}}^v_i \odot \widetilde{s}^v_{i-1} + \widetilde{\text{i}}^v_i \odot \widetilde{\text{g}}^v_i, \ \langle 3 \rangle \\
	s^v_{i-1}, \ \langle 4 \rangle
	\end{aligned}
	\right.
	;\ 
	s^v_i = \left\{
	\begin{aligned}
	\text{f}^v_i \odot s^v_{i-1} + \text{i}^v_i \odot \text{g}^v_i, \ \langle 3 \rangle \\
	s^v_{i-1}, \ \langle 4 \rangle
	\end{aligned}
	\right.,
	\end{aligned}
	\end{equation}
	where $\langle 3\rangle \langle 4\rangle$ respectively denote the conditions: $g_i=1,g_i=0$.
	We call the operation under $\langle 3\rangle$ \emph{update}, and it under $\langle 4\rangle$ \emph{copy}: When there is a click at $d_i$, then there should be the contextual information passed from the CTR layer, and thus \emph{update} operation will update the states at $d_i$.
	Otherwise (i.e., there is no click at $d_i$), then \emph{copy} operation will simply copy the states and hidden states of the previous timestep without fusing any information.
	The hidden states $\widetilde{h}^v_i$ and $h^v_i$ can be computed via 
	\begin{equation}
	\label{eqn:hiddencvr}
	\begin{aligned}
	\widetilde{h}^v_i = \left\{
	\begin{aligned}
	\widetilde{\text{o}}^v_i \odot \text{tanh}(\widetilde{s}^v_i), \ \langle 3 \rangle \\
	\widetilde{h}^v_{i-1},  \ \langle 4 \rangle\\
	\end{aligned}
	\right.
	;\
	h^v_i = \left\{
	\begin{aligned}
	\text{o}^v_i \odot \text{tanh}(s^v_i), \ \langle 3 \rangle \\
	h^v_{i-1},  \ \langle 4 \rangle\\
	\end{aligned}
	\right.
	\end{aligned}
	.
	\end{equation}
	
	

	\minisection{Inter-Layer Mechanism}
    $\text{i}_i$, $\text{o}_i$, $\text{g}_i$ are designed to encode the top-down contextual formation.
    Formally, for each item $d_i$, their values are updated as follows:
	\begin{equation}
	\label{eqn:cellgate}
	\begin{aligned}
	\text{f}_i &= \text{sigmoid } (\text{MLP}(\text{s}_i)), \ 
	\text{i}_i = \text{sigmoid } (\text{MLP}(\text{s}_i)), \\
	\text{o}_i &= \text{sigmoid } (\text{MLP}(\text{s}_i)), \ 
    \text{g}_i = \text{tanh } (\text{MLP}(\text{s}_i)), \\
	\end{aligned}
	\end{equation}
	where $\text{s}_i$ is the top-down state, computed as
	\begin{equation}
	\label{eqn:topdownstate}
	\begin{aligned}
	\text{s}^c_i &= [(1-g_{i-1}) \cdot U^c_{i-1} \cdot h^c_{i-1} + g_{i-1} \cdot U^r_{i-1} \cdot h^v_{i-1}],\\
	\text{s}^v_i &= [U^v_{i-1} \cdot h^v_{i-1} + g_i \cdot W^r_i \cdot h^c_i].
	\end{aligned}
	\end{equation}
	where $U^c_{i-1}$, $U^r_{i-1}$, $U^v_{i-1}$ and $W^r_i$ are trainable weights.
	
	Note that the above operations in Eqs.~(\ref{eqn:hiddenctr}), (\ref{eqn:hiddencvr}), (\ref{eqn:cellgate}), (\ref{eqn:topdownstate}) implicitly force the CVR layer to absorb the summary information from the CTR layer according to the top-down contexts. 
	Also, these operations are not assigned with a fixed update intervals, and thus can be adaptively adjusted corresponding to different contexts.
	
	\subsection{Loss Function}
	\label{sec:lossbias}
	Considering that both click and conversion signals are binary, we adopt binary cross entropy (BCE) loss as 
	\begin{equation}
	\label{eqn:all}
	 L = L_c + \alpha \cdot L_v \text{ where }
	\end{equation}
	\begin{equation}
	\label{eqn:biasloss}
	L_b
	= -\sum_{(b,\bm{x}_q)\in\mathcal{D}_q}\left(b\cdot \log P(b|\bm{x}_q) + (1-b)\cdot \log(1-P(b|\bm{x}_q))\right),
	\end{equation}
	where $b$ can denote either click $c$ or conversion $v$.
	
	For (biased) LTR, for each item $d_i$, the estimation of its click is $P(c_i=1)=P(r^c_i=1)=h^c_i$, and its conversion is $P(v_i=1)=P(r^v_i=1)\cdot P(c_i=1)=h^c_i\cdot h^v_i$. 
	
    \section{HEROES for Unbiased LTR}
    \label{sec:unbias}
    In this section, we extend the HEROES architecture into unbiased LTR (i.e., using Eq.~(\ref{eqn:prior})). 
    We first describe how to use the HEROES to model the user's multiple behavior through the entire space behavior path, and then present the corresponding loss function. 
    
    
    \subsection{Behavior Modelling on the Entire Space}
    Note that the HEROES introduced in Section~\ref{sec:model} that does not explicitly model the behavior path ``observation$\rightarrow$click''.
    However, as stated in Section~\ref{sec:back}, unbiased LTR requires the HEROES to use Eq.~(\ref{eqn:prior}) and learn the multiple user behaviors through the entire space behavior path ``observation$\rightarrow$click$\rightarrow$conversion''.
    
    To this end, we introduce the survival analysis technique \citep{jin2020deep,ren2019deep} to simultaneously estimate the user's observations, clicks, conversions.
    The main assumption of the survival analysis technique is that a patient will keep \emph{survival} until she \emph{leaves} the hospital or meets \emph{death}, which follows the path ``survival$\rightarrow$death''.
    
    Analogously, we can use it to formulate the user's multiple behaviors, which are similar but hierarchical:
	In the CTR layer, a user will keep \emph{observing} until she \emph{leaves} due to the lost interest or \emph{clicks} an item to check its detailed information, which follows the path ``observation $\rightarrow$ click''; where
	in the CVR layer, a user will keep \emph{clicking} to search for a worthwhile item until she \emph{leaves} due to mismatching between the user requirement and the current item or \emph{purchases} an item due to success in finding a favorite item, which follows the path ``click $\rightarrow$ conversion''.
    
    Based on the analysis above, we can define the probability density function (P.D.F.) of a user behavior occurring at $i$-th item $d_i$ as 
	\begin{equation}
	\label{eqn:pdf}
	\begin{aligned}
	P(c_i=1) = P(z^c=i), \ P(v_i=1) = P(z^v=i),
	\end{aligned}
	\end{equation}
	where $z^c$ and $z^v$ represent the click and conversion behaviors respectively; and $z=i$ means that the behavior occurs in $d_i$ and $z \geq i$ means that the behavior occurs after $d_i$.

	From the analogy between ``survival $\rightarrow$ death'' and ``observation $\rightarrow$ click'', ``click $\rightarrow$ conversion'', we can find that when studying the click behavior $z^c$ in the CTR layer, the CDF in this case (i.e., $P(z^c \geq i)$) denotes the observation probability, since a user will keep browsing until she finds an interesting item and clicks to check details.
	Similarly, if studying the conversion behavior $z^v$ in the CVR layer, the CDF here (i.e., $P(z^v \geq i)$) denotes the click probability since a user will keep clicking the items until she eventually purchases a favored one.
	Thus, we have
	\begin{equation}
	\label{eqn:cdf}
	\begin{aligned}
	P(o_i=1) = P(z^c \geq i), \ P(c_i=1) = P(z^v \geq i).
	\end{aligned}
	\end{equation}
	We then can derive the prior-click relevance $r^c_i$ and the post-click relevance $r^v_i$ by the conditional click probability $h^c_i$ and the conditional conversion probability $h^v_i$, which can be formulated as
	\begin{equation}
	\label{eqn:clickrelevance}
	h^c_i \coloneqq P(r^c_i = 1) = \frac{P(c_i=1)}{P(o_i=1)} = \frac{P(z^c=i)}{P(z^c \geq i)},
	\end{equation}
	\begin{equation}
	\label{eqn:conversionrelevance}
	h^v_i \coloneqq P(r^v_i = 1) = \frac{P(v_i=1)}{P(c_i=1)} = \frac{P(z^v=i)}{P(z^v \geq i)},
	\end{equation}
	which also indicates the probability that the click behavior $z^c$ (or conversion behavior $z^v$) lies at $d_i$ given the condition that $z^c$ (or $z^v$) is larger than the last observation (or click) boundary.
	For each layer, according to Eqs.~(\ref{eqn:cdf}), (\ref{eqn:clickrelevance}) and (\ref{eqn:conversionrelevance}), we can derive that 
	\begin{equation}
	\label{eqn:eqncdf}
	\begin{aligned}
	&P(z \geq i| \bm{x}_q; \theta) = P(z \neq 1, z \neq 2, \dots , z \neq i-1 | \bm{x}_q; \theta)\\
	=& \prod_{\tau:\tau < i} [1-P(z = \tau | z \geq \tau,\bm{x}_q; \theta)] = \prod_{\tau:\tau < i} (1-h_\tau).
	\end{aligned}
	\end{equation}
	Here, we use the probability chain rule to calculate the observation probability $P(i \leq z^c)$ (or the click probability $P(i \leq z^v)$) at $d_i$ through multiplying the conditional unclick probability $1-h^c_\tau$ (or the conditional unconversion probability $1-h^v_\tau$).
	Combining Eqs.~(\ref{eqn:clickrelevance}), (\ref{eqn:conversionrelevance}) and (\ref{eqn:eqncdf}) together, we have
	\begin{equation}
	\label{eqn:eqnh}
	P(z = i | \bm{x}_q; \theta) = h_{i} \cdot \prod_{\tau:\tau<i} (1 - h_\tau).
	\end{equation}
	Based on above description, we can express all the user's behaviors including observation $P(o_i=1)$, click $P(c_i=1)$, conversion $P(v_i=1)$ on $d_i$ by $h_i$.
	
	One straightforward approach to estimate $h_i$ is directly applying Eq.~(\ref{eqn:rnn}).
	However, it might cause \emph{incompatible issue}: Eq.~(\ref{eqn:rnn}) uses the Hawkes process to model the user's multiple behaviors (i.e., Eq.~(\ref{eqn:hawke})), while Eq.~(\ref{eqn:clickrelevance}) is derived from the survival analysis.
	Therefore, another way is to first tweak the HEROES unit by replacing $s^c(t_{i-1})$ by $s^c_{i-1}$ in Eq.~(\ref{eqn:cellctr}), and then employ HEROES (i.e., Eq.~(\ref{eqn:rnn})) to obtain $h_i$. 
	We will later establish an empirical investigation of the performance of these two approaches.

    \subsection{Loss Function on the Entire Space}
    \label{sec:lossunbias}
	
	
	The first type of loss is based on the PDF. 
	In the CTR and CVR layers, we aim to minimize the negative log-likelihood of the clicked or purchased item $d_j$ (i.e., $z=j$) as
	\begin{equation}
	\label{eqn:zc}
	\begin{aligned}
	L_\mathtt{pdf} & = - \text{log} \prod_{(\bm{x}_q, z) \in \mathcal{D}_{q}} P(z = j | \bm{x}_q; \theta) \\
	& = - \sum_{(\bm{x}_q, z) \in \mathcal{D}_q} [\text{log} \ h_j + \sum_{\tau: \tau<j} \text{log}(1-h_\tau)] ,
	\end{aligned}
	\end{equation}
	
	The second type of loss is based on the CDF.
	Let $I$ denote the length of the item list.
	There are two cases:
	(i) for those lists where there is a click occurring in the CTR layer or a conversion occurring in the CVR layer (i.e., $z \leq I$), we have
	\begin{equation}
	\label{eqn:click}
	\begin{aligned}
	L_\mathtt{occur} & = - \text{log} \prod_{(\bm{x}_q, I) \in \mathcal{D}_{\mathtt{occur}}} P(I \geq z|\bm{x}_q;\theta) \\
	& = - \sum_{(\bm{x}_q,I) \in \mathcal{D}_{\mathtt{occur}}} \text{log} \ [1 - \prod_{\tau: \tau \leq I} (1 - h_\tau)],
	\end{aligned}
	\end{equation}
	where $\mathcal{D}_{\mathtt{occur}}$ is the dataset of the above lists.
For the other lists, there is no click occurring in the CTR layer or no conversion occurring in the CVR layer (i.e., $z > I$), we have  
	\begin{equation}
	\label{eqn:nonclick}
	\begin{aligned}
	L_{\mathtt{non-occur}} & = - \text{log} \prod_{(\bm{x}_q,I) \in \mathcal{D}_{\mathtt{non-occur}}} P(z > I| \bm{x}_q; \theta)\\
	& = - \sum_{(\bm{x}_q,I) \in \mathcal{D}_{\mathtt{non-occur}}} \sum_{\tau: \tau \leq I} \text{log} \ (1-h_\tau),
	\end{aligned}
	\end{equation}
	where $\mathcal{D}_{\mathtt{non-occur}}$ is the dataset of the above lists, and $\mathcal{D}_q=\mathcal{D}_{\mathtt{occur}} \cup \mathcal{D}_{\mathtt{non-occur}}$.

    Combining all the objective functions (i.e., C.D.F. and P.D.F. losses), our goal is to minimize the negative log-likelihood over all the data samples as
	\begin{equation}
	\label{eqn:unbiasloss}
	L_b =  L_\mathtt{pdf} + \beta \cdot L_\mathtt{cdf} \text{ where } L_\mathtt{cdf} = L_{\mathtt{non-occur}} + L_{\mathtt{occur}},
	\end{equation}
	where all these losses are computed over the behavior $z^b$ (where $b$ can either denote $c$ or $v$), and the hyper-parameter $\beta$ balances P.D.F. and C.D.F. losses at the same level to stabilize the model training.
	The overall loss $L$ can be derived by applying Eq.~(\ref{eqn:unbiasloss}) into Eq.~(\ref{eqn:all}).
	
	For unbiased LTR, for each item $d_i$, according to Eq.~(\ref{eqn:eqnh}), the estimation of its click is $P(c_i=1)=P(z^c=i)=h^c_i\cdot\prod_{\tau:\tau<i}(1-h^c_\tau)$, and it conversion is $P(v_i=1)=P(z^v=i)=h^v_i\cdot\prod_{\tau:\tau<i}(1-h^v_\tau)$.
	
	\subsection{Overall Algorithm}
	We show the overall algorithm in Algorithm~\ref{algo:framework}.
	It is not hard to see that the main components of HEROES are the hierarchical recurrent neural network.
	Let $I_\text{max}$ denote the maximal length of document lists.
	Then, the calculation of each HEROES unit $f_\theta$ will run for maximal $2 I_\text{max}$ times.
	We assume the average case time performance of $f_\theta$ is $O(C)$.
	The subsequent calculation is to obtain the multiplication results of $h_i$ or $1-h_i$ to form the losses, whose complexity is $O(I_\text{max})$.
	Then, the overall time complexity is $O(2 C  I_\text{max})+O(I_\text{max})=O(2 C  I_\text{max})$.
	
	Note that we use \emph{behavioral relevance} $r_i$ for training to mine its latent \emph{inherent relevance} $\widetilde{r}_i$ for inference.
	Take the HEROES for biased LTR as an example.
	During training, we compute the loss in Eq.~(\ref{eqn:biasloss}) using $P(c_i=1)=P(r^c_i=1)=h^c_i$ for the CTR prediction and $P(v_i=1)=P(c_i=1)\cdot P(r^v=1)=h^c_i\cdot h^v_i$ for the CVR prediction.
    During inference, we use $P(c_i=1)=P(\widetilde{r}^c_i=1)=\widetilde{h}^c_i$ for the CTR estimation and $P(v_i=1)=P(c_i=1)\cdot P(\widetilde{r}^v_i=1)=\widehat{h}^c_i \cdot\widetilde{h}^v_i$ for the CVR estimation.
    Comparing to the existing pipeline (e.g., \citep{zhao2019recommending}) using $h^c_i$ and $h^c_i\cdot h^v_i$ for both training and inference, the proposed pipeline is able to encode the effect from the user's previous behaviors and recover the true inherent relevance.
	
		\begin{algorithm}[t!]
		\caption{HEROES}
		\label{algo:framework}
		\begin{algorithmic}[1]
			\STATE Initialize all parameters.
			\REPEAT
			\STATE Randomly sample a batch $\mathcal{B}$ from $\mathcal{D}$.
			\FOR {each item $d_i$ with $\mathcal{D}_q$ in $\mathcal{B}$}
			\STATE Calculate $h^c_i$, $h^v_i$, $\widetilde{h}^c_i$, $\widetilde{h}^v_i$ using Eq.~(\ref{eqn:rnn}).
			\ENDFOR
			\STATE Compute loss $L$ using $h^c_i$, $h^v_i$ by Eq.~(\ref{eqn:all}).
			\label{line:all}
			\COMMENT{Training}
			\STATE Update parameters $\theta$ by minimizing $L$.
			\STATE Generate ranking according to $\widetilde{h}^c_i$, $\widetilde{h}^v_i$.
			\label{line:inference}
			\COMMENT{Inference}
			\UNTIL convergence
		\end{algorithmic}
	\end{algorithm}

	\begin{table*}[t]
	\centering
	\caption{\small{Comparison of different multi-task models and sequential models on three industrial datasets. Results of both Click-Through Rate (CTR) and Conversion Rate (CVR) are reported. Bold values are the best in each column, while the second best values are underlined.
	* indicates $p < 0.001$ in significance tests compared to the best baseline.}}
	\vspace{-4mm}
	\resizebox{0.85\textwidth}{!}{
		\begin{tabular}{@{\extracolsep{4pt}}ccccccccccc}
			\toprule
			\multirow{2}{*}{Ranker} & \multirow{2}{*}{Task} & \multicolumn{3}{c}{Criteo} & \multicolumn{3}{c}{Taobao E-Commerce} & \multicolumn{3}{c}{Diantao Live Broadcast} \\
			\cmidrule{3-5}
			\cmidrule{6-8}
			\cmidrule{9-11}
			{} & {} & AUC & LogLoss & NDCG & AUC & LogLoss & NDCG & AUC & LogLoss & NDCG \\
			\midrule
			\multirow{2}{*}{DUPN} & CVR & 
			0.9505 & 0.1137 & 0.7348 & 
			\underline{0.6747} & 0.5194 & 0.6843 & 
			0.8232 & \underline{0.2345} & 0.7522 \\ 
			{} & CTR & 
			0.7410 & 0.5863 & 0.7526 & 
			0.5777 & 0.7215 & 0.4576 & 
			0.7156 & 0.6032 & 0.7009\\
			\midrule
			\multirow{2}{*}{ESMM} & CVR & 
			0.8750 & 0.4466 & 0.7194 & 
			0.6443 & 0.6330 & 0.6490 &
			0.7046 & 0.2743 & 0.6697\\
			{} & CTR & 
			0.6476 & 0.6511 & 0.7460 & 
			0.5410 & 0.7591 & 0.4166 & 
			0.6664 & 0.6577 & 0.6601\\
			\midrule
			\multirow{2}{*}{ESM$^2$} & CVR & 
			0.8798 & 0.4360 & 0.7235 & 
			0.6453 & 0.6376 & 0.6471 & 
			0.7039 & 0.2756 & 0.6688\\
			{} & CTR & 
			0.6740 & 0.6370 & 0.7496 & 
			0.5437 & 0.7573 & 0.4170 & 
			0.6742 & 0.6512 & 0.6608\\
			\midrule
			\multirow{2}{*}{MMoE} & CVR & 
			0.8817 & 0.4420 & 0.7182 & 
			0.6537 & 0.6267 & 0.6452 & 
			0.7283 & 0.2731 & 0.6653\\
			{} & CTR & 
			0.6779 & 0.6343 & 0.7540 & 
			0.5410 & 0.7463 & 0.4093 & 
			0.6770 & 0.6513 & 0.6618\\
			\midrule
            \multirow{2}{*}{DRSR} & CVR & 
			0.9468 & 0.1366 & 0.7644 & 
			0.6723 & 0.5156 & 0.6892 &
			0.8140 & 0.2546 & 0.7697\\
			{} & CTR & 
			0.7452 & 0.5837 & 0.7687 & 
			0.5759 & 0.7171 & \underline{0.4578} & 
			0.6985 & 0.6103 & 0.7053\\
			\midrule
			\multirow{2}{*}{RRN} & CVR & 
			0.9564 & 0.1169 & 0.7739 & 
			0.6732 & 0.5061 & 0.6890 &
			0.8156 & 0.2698 & 0.7421\\
			{} & CTR & 
			0.7496 & 0.5797 & 0.7706 & 
			0.5766 & \underline{0.7075} & 0.4575 & 
			0.6926 & 0.6019 & 0.6928\\
			\midrule
			\multirow{2}{*}{NARM} & CVR & 
			0.9524 & 0.1172 & 0.7644 & 
			0.6733 & 0.5160 & \underline{0.6893} &
			0.8234 & 0.2595 & 0.7612\\
			{} & CTR & 
			0.7511 & 0.5810 & 0.7724 & 
			0.5764 & 0.7186 & 0.4576 & 
			0.7082 & 0.5958 & 0.7012\\
			\midrule
			\multirow{2}{*}{STAMP} & CVR & 
			0.9406 & 0.1209 & 0.8014 & 
			0.6668 & 0.5210 & 0.6892 &
			0.8467 & 0.2465 & 0.7689\\
			{} & CTR & 
			0.7391 & 0.5929 & 0.7702 & 
			0.5748 & 0.7235 & 0.4575 & 
			0.7123 & \underline{0.5940} & 0.7070\\
			\midrule
			\multirow{2}{*}{Time-LSTM} & CVR & 
			\underline{0.9622} & 0.1132 & \underline{0.7979} & 
			0.6745 & 0.5169 & 0.6889 &
			\underline{0.8540} & 0.2412 & \underline{0.7787}\\
			{} & CTR & 
			\underline{0.7602} & \underline{0.5703} & \underline{0.7738} & 
			\underline{0.5776} & 0.7192 & 0.4576 & 
			\underline{0.7195} & 0.6040 & \underline{0.7124}\\
			\midrule
			\multirow{2}{*}{LSTM} & CVR & 
			0.8429 & 0.4841 & 0.6629 & 
			0.6721 & \underline{0.4783} & 0.6885 &
			0.7124 & 0.2736 & 0.7475\\
			{} & CTR & 
			0.6032 & 0.6042 & 0.7503 & 
			0.5749 & 0.7222 & 0.4493 & 
			0.6633 & 0.6542 & 0.6792\\
			\midrule
			\multirow{2}{*}{NHP} & CVR & 
			0.9533 & \underline{0.1127} & 0.7682 & 
			0.6743 & 0.4914 & \underline{0.6893} &
			0.8267 & 0.2535 & 0.7622\\
			{} & CTR & 
			0.7428 & 0.5816 & 0.7656 & 
			0.5773 & 0.7214 & 0.4576 & 
			0.7033 & 0.6042 & 0.7068\\
			\bottomrule
			\toprule
			\multirow{2}{*}{$\text{HEROES}^-_\mathtt{intra}$} & CVR & 
			0.8801 & 0.4270 & 0.7327 & 
			0.6917 & 0.5209 & 0.6998 & 
			0.8045 & 0.2675 & 0.7712\\
			{} & CTR & 
			0.6764 & 0.6612 & 0.7521 & 
			0.5483 & 0.7174 & 0.4682 & 
			0.7091 & 0.5976 & 0.7135\\
			\midrule
            \multirow{2}{*}{$\text{HEROES}^-_\mathtt{inter}$} & CVR & 
			0.9682 & 0.1152 & 0.7832 & 
			0.6932 & 0.4918 & 0.7082 &
			0.8346 & 0.2225 & 0.7883\\
			{} & CTR & 
			0.7632 & 0.5721 & 0.7882 & 
			0.5927 & 0.7032 & 0.4721 & 
			0.7138 & 0.6021 & 0.7123\\
			\midrule
			\multirow{2}{*}{$\text{HEROES}^-_\mathtt{unit}$} & CVR & 
			0.9705 & 0.1016 & 0.8348 & 
			0.7402 & 0.4366 & 0.7106 &
			0.8601 & 0.2350 & 0.7810\\
			{} & CTR & 
			0.7787 & 0.5483 & 0.7832 & 
			0.5920 & 0.7084 & 0.4701 & 
			0.7412 & 0.5942 & 0.7111\\
			\midrule
            \multirow{2}{*}{\textbf{HEROES}} & CVR & 
			\textbf{0.9759}$^*$ & \textbf{0.0975}$^*$ & \textbf{0.8551}$^*$ & 
			\textbf{0.7503}$^*$ & \textbf{0.3519}$^*$ & \textbf{0.7137}$^*$ &
			\textbf{0.8649}$^*$ & \textbf{0.2203}$^*$ & \textbf{0.7893}$^*$\\
			{} & CTR & 
			\textbf{0.7870}$^*$ & \textbf{0.5400}$^*$ & \textbf{0.7913}$^*$ & 
			\textbf{0.5953}$^*$ & \textbf{0.7024}$^*$ & \textbf{0.4727}$^*$ & 
			\textbf{0.7492}$^*$ & \textbf{0.5893}$^*$ & \textbf{0.7166}$^*$\\
			\bottomrule
		\end{tabular}
	}
	\label{tab:res}
	\vspace{-2mm}
\end{table*}	

\begin{table}[t]
	\centering
	\caption{\small{Comparison of unbiased LTR and biased LTR version of HEROES under click generation model PBM.}}
	\vspace{-3mm}
	\resizebox{0.95\linewidth}{!}{
		\begin{tabular}{@{\extracolsep{4pt}}ccccc}
			\toprule
			\multirow{2}{*}{Ranker} & \multirow{2}{*}{Task} & \multicolumn{3}{c}{Taobao E-Commerce (PBM)} \\
			\cmidrule{3-5}
			{} & {} & AUC & LogLoss & NDCG\\
			\midrule
			\multirow{2}{*}{Relevance Data (HEROES)} & CVR & 
			0.7503 & 0.3519 & 0.7137 \\
			{} & CTR & 
			0.5953 & 0.7024 & 0.4727 \\
			\midrule
			\multirow{2}{*}{\textbf{HEROES$^+$}} & CVR & 
			0.7442 & 0.3674 & 0.7064 \\
			{} & CTR & 
			0.5735 & 0.7206 & 0.4567 \\
			\midrule
			\multirow{2}{*}{\textbf{HEROES$^+_\mathtt{comb}$}} & CVR & 
			0.7463 & 0.3638 & 0.7110 \\
			{} & CTR & 
			0.5738 & 0.7202 & 0.4521 \\
			\midrule
			\multirow{2}{*}{Click Data (HEROES)} & CVR & 
			0.7412 & 0.3746 & 0.7024 \\
			{} & CTR & 
			0.5643 & 0.7563 & 0.4284 \\
			\bottomrule
		\end{tabular}
	}
	\vspace{-2mm}
	\label{tab:bias}
\end{table}

	\section{Experiment}
    \subsection{Dataset Description and Evaluation Flow}
    \label{app:dataset}
    We use three large-scale real-world datasets for the evaluations, where the first two are public benchmark datasets and the last one is created by our own:
    \begin{itemize}[topsep = 3pt,leftmargin =5pt]
		\item \textbf{Criteo dataset}\footnote{\url{https://ailab.criteo.com/ressources/}} is formed of Criteo live traffic data in a period of 30 days.
		It consists of more than 5.5 million impressions with 2.3 million clicks and 438 thousand conversions.
		Since the query signal is not available, following \citep{ren2018learning}, we incorporate the user ID and conversion ID to divide the full dataset into several sequential data.
		In this way, we can obtain 2.2 million item lists (i.e., queries). 
		\item \textbf{Taobao E-Commerce dataset}\footnote{\url{https://tianchi.aliyun.com/datalab/dataSet.html?dataId=408}} is collected from the traffic logs of Taobao's recommender system.
		It contains the logs of 444 thousand users browsing 85 million items under 1,614 thousand queries.
		In these queries, there are sequential user behaviors, including 3,317 thousand click labels and 17 thousand conversion signals.
		\item \textbf{Diantao Live Broadcast Recommendation dataset} is collected from the user interaction logs of Diantao App which shares the same database of users and anchors with Taobao App.
		It contains more than 44 million logs of 905 thousand users' browsing histories over 527 thousand items in 9,305 thousand queries.
		Features of the user include age, gender, city, etc., and features of the document include title, time, etc.   
		In each query, we regard the items whose playtime more than $7$s as the clicked ones.
		And we further treat the clicked items that are liked or commented on by the user as the purchased ones.
	\end{itemize}
	For each dataset, we split the users' sequential historical records by queries into training/validation/test datasets at a ratio of 6:2:2.
	For fair comparison, we do not input the user's behaviors (i.e., $y$) into $f_\gamma(\cdot)$ in Eq.~(\ref{eqn:st}).
	We train each model with training queries and evaluate its performance with new queries.
	In order to evaluate the above methods, for each method, we choose Area under the ROC Curve (AUC), LogLoss, Normalized Distributed Cumulative Gain (NDCG) as evaluation measures.
	Specifically, we calculate NDCG over the whole ranking list. 
	Namely, we compute NDCG@K where K is the length of the ranking list.
	

    \subsection{Experimental Configuration}
    \label{app:experiment}
    As we conduct the experiments in terms of the CTR and CVR predictions, the most related existing baselines are those originally proposed for multi-task learning, including \textbf{DUPN} \citep{ni2018perceive}, \textbf{ESMM} \citep{ma2018entire}, \textbf{ESM$^2$} \citep{wen2019entire}, \textbf{MMoE} \citep{ma2018modeling}.
    We further extend the existing methods originally designed for the CTR prediction into these multi-task setting by using two individual models independently learning and predicting for the CTR and CVR tasks respectively.
    As one of the main contributions in the paper is to design a novel sequential unit, thus, we mainly include the following sequential models here:
    \textbf{LSTM} \citep{hochreiter1997long}, \textbf{RRN} \citep{wu2017recurrent},  \textbf{NARM} \citep{li2017neural}, \textbf{NARM} \citep{li2017neural}, \textbf{STAMP} \citep{liu2018stamp}, \textbf{DRSR} \citep{jin2020deep}, \textbf{NHP} \citep{mei2016neural}.

    \begin{figure*}[t]
	\centering
	\includegraphics[width=1.00\linewidth]{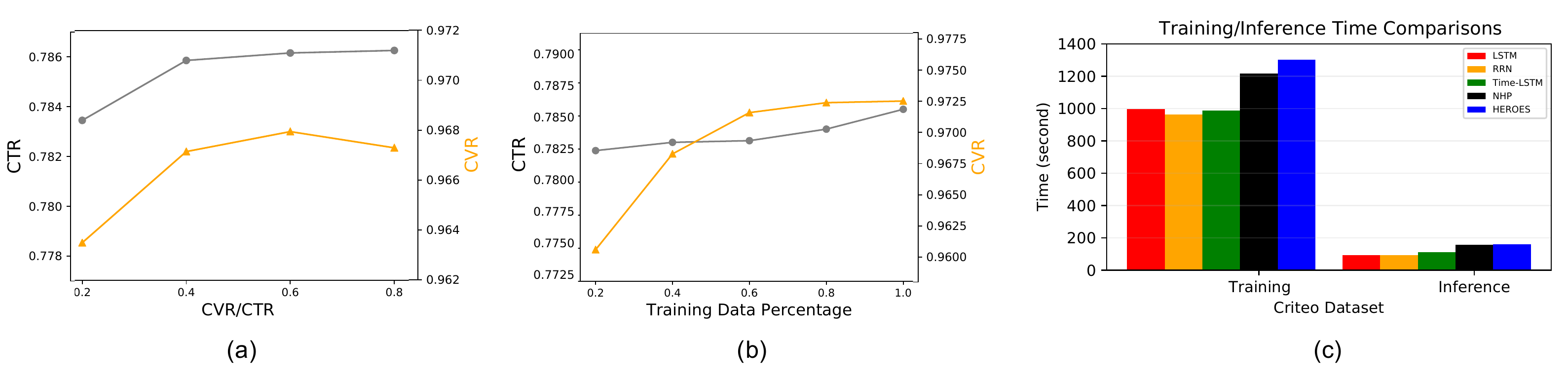}
	\vspace{-9mm}
	\caption{\small{(a) Performance change of HEROES against the ratio of CVR and CTR loss weights. (b) Performance change of HEROES against click data with different amounts of training data. (c) Training/inference time comparisons of HEROES against the sequential models.}}
	\label{fig:biascurve}
	\vspace{-4mm}
    \end{figure*}

	Besides these baselines, we introduce several variants of HEROES as ablations.
	More specifically, in order to further investigate the effect from each component, we design the following three variants:
	\begin{itemize}[topsep = 3pt,leftmargin =5pt]
    \item $\textbf{HEROES}^-_\mathtt{intra}$:
    we adopt a MLP, instead of our sequential model to model the correlations within each layer in Section~\ref{sec:architecture}.
    \item $\textbf{HEROES}^-_\mathtt{inter}$:
    we train the CTR and CVR layers independently without modeling the correlations cross two layers in Section~\ref{sec:architecture}.
	\item $\textbf{HEROES}^-_\mathtt{unit}$:
	we adopt a standard LSTM unit instead of one introduced in Section~\ref{subsec:unit}.
	\end{itemize}
	For further evaluation in the different settings, we clarify the use of HEROES in the context of biased LTR and unbiased LTR as 
	\begin{itemize}[topsep = 3pt,leftmargin =5pt]
	\item \textbf{HEROES} is the HEROES for biased LTR setting, where we follow
	Section~\ref{sec:lossbias}
	to produce the loss and the CTR and CVR predictions.
	\item $\textbf{HEROES}^+_\mathtt{comb}$ is the HEROES with unbiased LTR setting, where we follow Section~\ref{sec:lossunbias} to produce the loss and the CTR and CVR predictions.
	\item $\textbf{HEROES}^+$ is a variant of HEROES$^+_\mathtt{comb}$ where we tweak the HEROES unit by replacing $s^c(t_{i-1})$ with $s^c_{i-1}$ in Eq.~(\ref{eqn:cellctr}).
	\end{itemize}

	\subsection{Performance Comparison of CTR and CVR}

	Table~\ref{tab:res} summarizes the results.
	The major findings from our experiments are summarized as follows:
	\begin{itemize}[topsep = 3pt,leftmargin =5pt]
	\item The performance of HEROES is significantly better than the multi-task learning methods (including DUPN, ESMM, ESM$^2$, MMoE).
	One explanation is that although DUPN uses LSTM to encode the sequential data and MMoE incorporates mixture-of-expert structure, their performances are still limited by treating the multiple behaviors with the same time scales.
	\item  HEROES significantly outperforms the baselines (including LSTM, RRN, NARM, STAMP, Time-LSTM, DRSR, NHP).
	A potential reason is that although NHP uses the Hawkes process and Time-LSTM incorporates the time intervals into the LSTM, they do not consider the correlations among the user's multiple behaviors.
	\item Note that generally speaking, the conversions are harder to predict than the clicks, as the conversion signals are usually much sparser in the real-world scenarios.
	However, Table~\ref{tab:res} shows the opposite.
	We provide three possible explanations as follows.
	(i) For those multi-task learning models (e.g., HEREOS, DUPN), the CVR prediction can not only benefit from the conversion signals but also can promote the predictions of the click signals.
	(ii) There are usually less noises in the conversion signals than the click ones.
	(iii) For these three datasets, we observe that purchased items are often located at the end of the sequences, as users are likely to keep browsing until finding the favorite items. 
	\end{itemize}
	In order to deeply analyze the model design and its superiority, we conduct the following ablation studies on the Criteo dataset.

	\minisection{Effect of Architecture}
	We investigate the effect of our hierarchical architecture design by comparing our model to $\text{HEROES}^-_\mathtt{intra}$ and $\text{HEROES}^-_\mathtt{inter}$.
    As Table~\ref{tab:res} shows, MLP is not capable to encode the sequential patterns which verifies HEROES using sequential module for modelling intra-layer correlations.
	Also, we can see that the correlations across the layers can also benefit the CTR and CVR predictions, which verifies the necessity of building a up-down channel across the CTR and CVR layers.   
	
	\minisection{Effect of Unit}
	In order to specific the performance gain from HEROES unit design, we introduce HEROES$^-_\mathtt{unit}$, which keeps the hierarchical architecture but uses a standard LSTM unit.
	Results reported in Table~\ref{tab:res} demonstrates the improvements of the unit.
	
	\minisection{Effect of Loss Function Weight}
	In order to study the influence of choosing different weights for CTR and CVR losses, we assign different values to $\alpha$ in Eq.~(\ref{eqn:all}), where $\alpha=\alpha/1$ represents the ratio of CVR and CTR weights.
	Result depicted in Figure~\ref{fig:biascurve}(a) indicates both of these losses play an important role for the final performance.
	
	\minisection{Robustness Analysis}
	We investigate the robustness of HEROES with different amounts of training data.
    We first randomly select a subset of training data (i.e.,~20\% - 100\%) to generate click data and then use these datasets to train 
    HEROES model.  
    Figure~\ref{fig:biascurve}(b) shows that HEROES can still work well even with limited training data.
	
	\minisection{Complexity Analysis}
	We study the time complexity of HEROES against baseline methods LSTM, RRN, Time-LSTM, NHP which are the sequential models.
	From Figure~\ref{fig:biascurve}(c), we observe that during training, NHP and HEROES are the most time-consuming methods as they incorporate the Hawkes process in the recurrent networks; while their inference time are comparable to the other methods.

    \subsection{Performance Comparison of Unbiased LTR}
	In order to verify whether HEROES$^+$ can work in the context of the unbiased LTR, we follow the click data generation process from \cite{ai2018unbiaseda,hu2019unbiased,jin2020deep} to introduce the position bias, and conduct the experiment on Taobao E-Commerce data.
    First, we train a Rank SVM model using $1\%$ of the training data with relevance labels. 
    Next, we use the trained model to create an initial ranked list for each query. 
    Then, we simulate the user browsing processes and sample clicks from the initial list.
    Position-based Model (PBM) \citep{richardson2007predicting} simulates 
	the user browsing behavior based on the assumption that the bias of an item only depends on its position, which can be formulated as $P(o_i) = \rho_i^\tau$, where $\rho_i$ represents 
	position bias at position $i$ and $\tau \in [0, +\infty]$ is a parameter controlling the degree of position bias.
	The position bias $\rho_i$ is obtained from an eye-tracking experiment in \citep{joachims2005accurately} and the parameter $\tau$ is set as one by default. 
	It also assumes that a user decides to click a item $d_i$ according to the probability $P(c_i) = P(o_i) \cdot P(r_i)$.
	Since there is no typical investigation of the position bias on the behavior path ``click$\rightarrow$conversion'', we simply operate on the conversion signals by assigning $v=0$ for those items with no click (i.e., $c=0$) and keep the original values for those clicked items.
	We regard the initial lists as the relevance data (i.e., unbiased data) and generated lists as the click data (i.e., biased data).
	As Table~\ref{tab:bias} shows, our unbiased version, denoted as HEROES$^+$, outperforms the biased one, denoted as HEROES, which shows that HEROES$^+$ can mitigate the position bias.
	HEROES$^+$ and HEROES$^+_\mathtt{comb}$ achieve comparable results, which indicates that incorporating the Hawkes process into the survival analysis process would not cause much difference. 
	
\begin{figure}[t]
	\centering
	\includegraphics[width=1.00\linewidth]{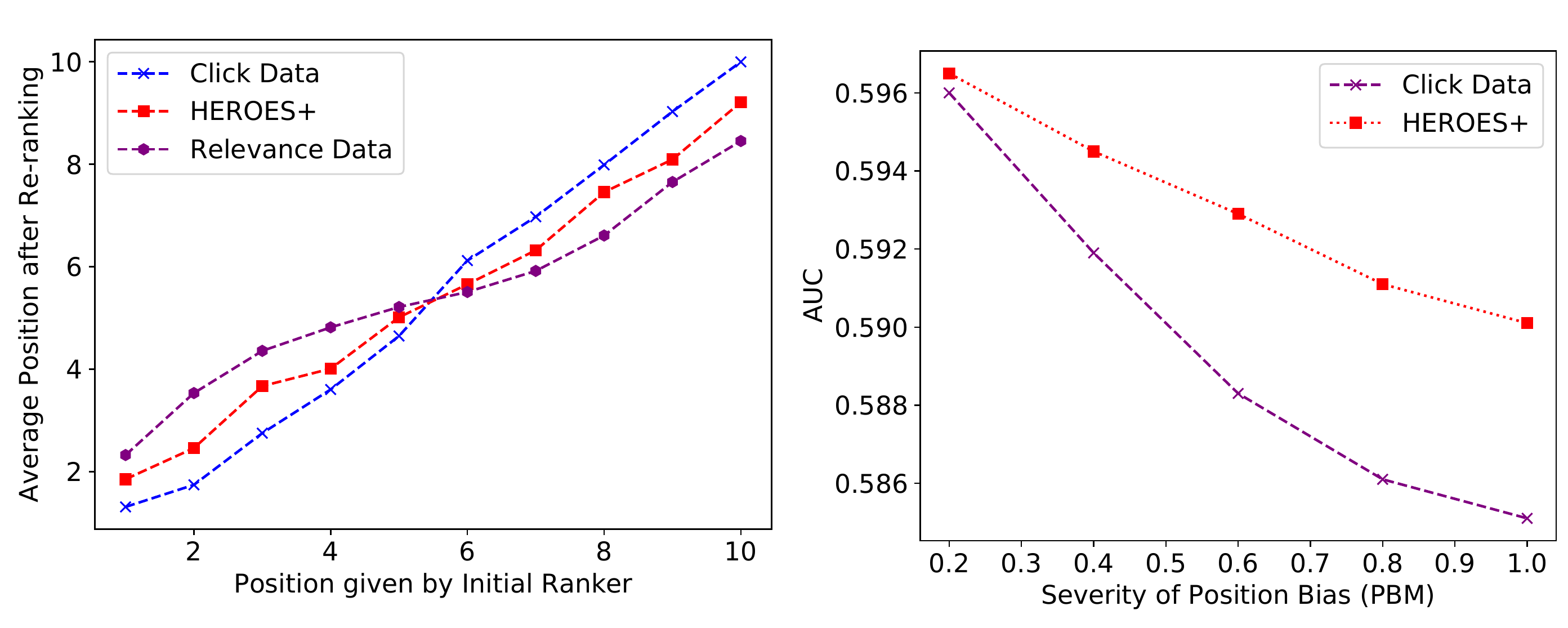}
	\vspace{-8.5mm}
	\caption{\small{(a) Average position after re-ranking of the item at each original position. (b) Performance change of HEROES$^+$ against click data with different degrees of position bias.}
	}
	\label{fig:unbiascurve}
	\vspace{-5.5mm}
    \end{figure}

	\minisection{Visualization Analysis}
	Here, we investigate whether the performance gain of HEROES$^+$ indeed owes to the reduction of position bias.
	We compare the ranking list given by the debiased ranker against the initial ranker.
	Specifically, we first identify the items at each position given by the initial ranker. 
	Then we calculate the average positions of the items at each original position after re-ranking. 
	We also calculate their average positions after re-ranking their relevance labels, which is regarded as the ground truth. 
	Ideally, the average positions produced by the debiasing methods should be close to the average position by relevance labels. 
	We summarize the results in Figure~\ref{fig:unbiascurve}(a). 
    The curve of HEROES$^+$ (in red with {\small ``$\small \blacksquare$''}) is the closest to the relevance label curve (in purple with mark `$\small \bullet$'), indicating that the performance enhancement of HEROES$^+$ is indeed due to effective debiasing.

	\minisection{Robustness Analysis}
	We evaluate the robustness of HEROES$^+$ under different degrees of position bias. 
	In the above experiments, we only test the performance of HEROES$^+$ with click data generated from click models with a given degree of position bias, i.e.,~$\tau=1$ in Taobao E-Commerce (PBM). 
	Therefore, we set the hyper-parameters for each click generation model to five values and examine whether HEROES$^+$ is still equally effective.
	Figure~\ref{fig:unbiascurve}(b) shows the AUC results as the degree of position bias increases; the results in terms of other measures follow similar trends. 

	\section{Related Work}
	
	There are a variety of user behaviors such as browsing, clicking (i.e., engagement behaviors), and rating, purchasing (i.e., satisfaction behaviors) \citep{zhao2019recommending}.
	There are mainly two directions of existing works on behavior awareness and multi-task information systems.
	One line of the research \citep{zhao2015improving,ma2018entire,wen2019entire,meng2020incorporating,wen2021hierarchically,xi2021modeling} is to investigate the behavior decomposition to learn and estimate multiple types of the user behaviors over all the samples.
	For example, \citet{wen2019entire} composes the post-click behaviors and develops a multi-task learning algorithm to combine these estimations to compute a final utility score for ranking.
	\citet{wen2021hierarchically} further incorporates the sequential behavior graph to encode the dependence among the user's multiple behaviors.
	The other direction of the research \citep{ma2018modeling,zhao2019recommending,wang2020m2grl,hadash2018rank,ni2018perceive,tang2020progressive,ding2021mssm,dai2021fishnet,zhao2019recommending,gao2019neural} aims to explicitly learn the task relationship and design an effective feature sharing technique.
	For instance, 
	\citet{ma2018modeling} and \citet{zhao2019recommending} adopts the Mixture-of-Experts model \citep{jacobs1991adaptive} to tradeoff between task-specific objectives and inter-task relationships.
	There is also a recently emerged direction \citep{wang2022escm,zhang2020large} studying the CTR and CVR estimations from a causal perspective.   
	However, all the above previous literature do not explicitly capture, or even are not aware of, the natural multi-scale characteristics of the user's multiple behaviors.
	Instead, our work establishes a hierarchical architecture which can incorporate the contextual information to automatically find a specific time scale for each behavior path (i.e., ``observation$\rightarrow$click'', ``click$\rightarrow$conversion'').

	\section{Conclusion and Future Work}
	In this paper, we propose a paradigm named HEROES, which can automatically discover the user's multi-scale browsing patterns to model the user's engagement and satisfaction behaviors.
	We design a novel recurrent unit to consider both excitation and discouragement from the contexts. 
	We further show that HEROES learn over the entire space behavior path ``observation$\rightarrow$click$\rightarrow$conversion'', and can be extended to unbiased LTR task.
	It would be interesting to investigate modelling more complex user behaviors (e.g., adding the items into the wish list or cart) in future work.
	
	\minisection{Acknowledgments} 
    This work was supported by Alibaba Group through Alibaba Research Intern Program.
    The Shanghai Jiao Tong University Team is supported by Shanghai Municipal Science and Technology Major Project (2021SHZDZX0102) and National Natural Science Foundation of China (62076161, 62177033).
    We would also like to thank Wu Wen Jun Honorary Doctoral Scholarship from AI Institute, Shanghai Jiao Tong University.

\clearpage
\bibliographystyle{ACM-Reference-Format}
\bibliography{heroes}


\begin{thebibliography}{37}


\ifx \showCODEN    \undefined \def \showCODEN     #1{\unskip}     \fi
\ifx \showDOI      \undefined \def \showDOI       #1{#1}\fi
\ifx \showISBNx    \undefined \def \showISBNx     #1{\unskip}     \fi
\ifx \showISBNxiii \undefined \def \showISBNxiii  #1{\unskip}     \fi
\ifx \showISSN     \undefined \def \showISSN      #1{\unskip}     \fi
\ifx \showLCCN     \undefined \def \showLCCN      #1{\unskip}     \fi
\ifx \shownote     \undefined \def \shownote      #1{#1}          \fi
\ifx \showarticletitle \undefined \def \showarticletitle #1{#1}   \fi
\ifx \showURL      \undefined \def \showURL       {\relax}        \fi
\providecommand\bibfield[2]{#2}
\providecommand\bibinfo[2]{#2}
\providecommand\natexlab[1]{#1}
\providecommand\showeprint[2][]{arXiv:#2}

\bibitem[\protect\citeauthoryear{Ai, Bi, Luo, Guo, and Croft}{Ai
  et~al\mbox{.}}{2018}]%
        {ai2018unbiaseda}
\bibfield{author}{\bibinfo{person}{Qingyao Ai}, \bibinfo{person}{Keping Bi},
  \bibinfo{person}{Cheng Luo}, \bibinfo{person}{Jiafeng Guo}, {and}
  \bibinfo{person}{W~Bruce Croft}.} \bibinfo{year}{2018}\natexlab{}.
\newblock \showarticletitle{Unbiased Learning to Rank with Unbiased Propensity
  Estimation}.
\newblock \bibinfo{journal}{\emph{SIGIR}} (\bibinfo{year}{2018}).
\newblock


\bibitem[\protect\citeauthoryear{Caruana}{Caruana}{1997}]%
        {caruana1997multitask}
\bibfield{author}{\bibinfo{person}{Rich Caruana}.}
  \bibinfo{year}{1997}\natexlab{}.
\newblock \showarticletitle{Multitask learning}.
\newblock \bibinfo{journal}{\emph{Machine learning}} \bibinfo{volume}{28},
  \bibinfo{number}{1} (\bibinfo{year}{1997}), \bibinfo{pages}{41--75}.
\newblock


\bibitem[\protect\citeauthoryear{Chung, Ahn, and Bengio}{Chung
  et~al\mbox{.}}{2016}]%
        {chung2016hierarchical}
\bibfield{author}{\bibinfo{person}{Junyoung Chung}, \bibinfo{person}{Sungjin
  Ahn}, {and} \bibinfo{person}{Yoshua Bengio}.}
  \bibinfo{year}{2016}\natexlab{}.
\newblock \showarticletitle{Hierarchical multiscale recurrent neural networks}.
  In \bibinfo{booktitle}{\emph{ICLR}}.
\newblock


\bibitem[\protect\citeauthoryear{Dai, Kong, Guo, and He}{Dai
  et~al\mbox{.}}{2021}]%
        {dai2021fishnet}
\bibfield{author}{\bibinfo{person}{Xin Dai}, \bibinfo{person}{Xiangnan Kong},
  \bibinfo{person}{Tian Guo}, {and} \bibinfo{person}{Xinlu He}.}
  \bibinfo{year}{2021}\natexlab{}.
\newblock \showarticletitle{FiShNet: Fine-Grained Filter Sharing for
  Resource-Efficient Multi-Task Learning}. In \bibinfo{booktitle}{\emph{CIKM}}.
\newblock


\bibitem[\protect\citeauthoryear{Ding, Dong, He, Cheng, Fu, Huan, Li, Yan,
  Zhang, Zhang, et~al\mbox{.}}{Ding et~al\mbox{.}}{2021}]%
        {ding2021mssm}
\bibfield{author}{\bibinfo{person}{Ke Ding}, \bibinfo{person}{Xin Dong},
  \bibinfo{person}{Yong He}, \bibinfo{person}{Lei Cheng},
  \bibinfo{person}{Chilin Fu}, \bibinfo{person}{Zhaoxin Huan},
  \bibinfo{person}{Hai Li}, \bibinfo{person}{Tan Yan}, \bibinfo{person}{Liang
  Zhang}, \bibinfo{person}{Xiaolu Zhang}, {et~al\mbox{.}}}
  \bibinfo{year}{2021}\natexlab{}.
\newblock \showarticletitle{MSSM: a multiple-level sparse sharing model for
  efficient multi-task learning}. In \bibinfo{booktitle}{\emph{SIGIR}}.
\newblock


\bibitem[\protect\citeauthoryear{Embrechts, Liniger, and Lin}{Embrechts
  et~al\mbox{.}}{2011}]%
        {embrechts2011multivariate}
\bibfield{author}{\bibinfo{person}{Paul Embrechts}, \bibinfo{person}{Thomas
  Liniger}, {and} \bibinfo{person}{Lu Lin}.} \bibinfo{year}{2011}\natexlab{}.
\newblock \showarticletitle{Multivariate Hawkes processes: an application to
  financial data}.
\newblock \bibinfo{journal}{\emph{Journal of Applied Probability}}
  \bibinfo{volume}{48}, \bibinfo{number}{A} (\bibinfo{year}{2011}),
  \bibinfo{pages}{367--378}.
\newblock


\bibitem[\protect\citeauthoryear{Gao, He, Gan, Chen, Feng, Li, Chua, and
  Jin}{Gao et~al\mbox{.}}{2019}]%
        {gao2019neural}
\bibfield{author}{\bibinfo{person}{Chen Gao}, \bibinfo{person}{Xiangnan He},
  \bibinfo{person}{Dahua Gan}, \bibinfo{person}{Xiangning Chen},
  \bibinfo{person}{Fuli Feng}, \bibinfo{person}{Yong Li},
  \bibinfo{person}{Tat-Seng Chua}, {and} \bibinfo{person}{Depeng Jin}.}
  \bibinfo{year}{2019}\natexlab{}.
\newblock \showarticletitle{Neural multi-task recommendation from
  multi-behavior data}. In \bibinfo{booktitle}{\emph{ICDE}}.
\newblock


\bibitem[\protect\citeauthoryear{Hadash, Shalom, and Osadchy}{Hadash
  et~al\mbox{.}}{2018}]%
        {hadash2018rank}
\bibfield{author}{\bibinfo{person}{Guy Hadash}, \bibinfo{person}{Oren~Sar
  Shalom}, {and} \bibinfo{person}{Rita Osadchy}.}
  \bibinfo{year}{2018}\natexlab{}.
\newblock \showarticletitle{Rank and rate: multi-task learning for recommender
  systems}. In \bibinfo{booktitle}{\emph{Recsys}}.
\newblock


\bibitem[\protect\citeauthoryear{Hawkes}{Hawkes}{1971}]%
        {hawkes1971spectra}
\bibfield{author}{\bibinfo{person}{Alan~G Hawkes}.}
  \bibinfo{year}{1971}\natexlab{}.
\newblock \showarticletitle{Spectra of some self-exciting and mutually exciting
  point processes}.
\newblock \bibinfo{journal}{\emph{Biometrika}} \bibinfo{volume}{58},
  \bibinfo{number}{1} (\bibinfo{year}{1971}), \bibinfo{pages}{83--90}.
\newblock


\bibitem[\protect\citeauthoryear{Hochreiter and Schmidhuber}{Hochreiter and
  Schmidhuber}{1997}]%
        {hochreiter1997long}
\bibfield{author}{\bibinfo{person}{Sepp Hochreiter} {and}
  \bibinfo{person}{J{\"u}rgen Schmidhuber}.} \bibinfo{year}{1997}\natexlab{}.
\newblock \showarticletitle{Long short-term memory}.
\newblock \bibinfo{journal}{\emph{Neural computation}} \bibinfo{volume}{9},
  \bibinfo{number}{8} (\bibinfo{year}{1997}), \bibinfo{pages}{1735--1780}.
\newblock


\bibitem[\protect\citeauthoryear{Hu, Wang, Peng, and Li}{Hu
  et~al\mbox{.}}{2019}]%
        {hu2019unbiased}
\bibfield{author}{\bibinfo{person}{Ziniu Hu}, \bibinfo{person}{Yang Wang},
  \bibinfo{person}{Qu Peng}, {and} \bibinfo{person}{Hang Li}.}
  \bibinfo{year}{2019}\natexlab{}.
\newblock \showarticletitle{Unbiased LambdaMART: An unbiased pairwise
  learning-to-rank algorithm}. In \bibinfo{booktitle}{\emph{WWW}}.
\newblock


\bibitem[\protect\citeauthoryear{Jacobs, Jordan, Nowlan, and Hinton}{Jacobs
  et~al\mbox{.}}{1991}]%
        {jacobs1991adaptive}
\bibfield{author}{\bibinfo{person}{Robert~A Jacobs}, \bibinfo{person}{Michael~I
  Jordan}, \bibinfo{person}{Steven~J Nowlan}, {and} \bibinfo{person}{Geoffrey~E
  Hinton}.} \bibinfo{year}{1991}\natexlab{}.
\newblock \showarticletitle{Adaptive mixtures of local experts}.
\newblock \bibinfo{journal}{\emph{Neural computation}} \bibinfo{volume}{3},
  \bibinfo{number}{1} (\bibinfo{year}{1991}), \bibinfo{pages}{79--87}.
\newblock


\bibitem[\protect\citeauthoryear{Jin, Fang, Zhang, Ren, Zhou, Xu, Yu, Wang,
  Zhu, and Gai}{Jin et~al\mbox{.}}{2020}]%
        {jin2020deep}
\bibfield{author}{\bibinfo{person}{Jiarui Jin}, \bibinfo{person}{Yuchen Fang},
  \bibinfo{person}{Weinan Zhang}, \bibinfo{person}{Kan Ren},
  \bibinfo{person}{Guorui Zhou}, \bibinfo{person}{Jian Xu},
  \bibinfo{person}{Yong Yu}, \bibinfo{person}{Jun Wang},
  \bibinfo{person}{Xiaoqiang Zhu}, {and} \bibinfo{person}{Kun Gai}.}
  \bibinfo{year}{2020}\natexlab{}.
\newblock \showarticletitle{A Deep Recurrent Survival Model for Unbiased
  Ranking}. In \bibinfo{booktitle}{\emph{SIGIR}}.
\newblock


\bibitem[\protect\citeauthoryear{Joachims, Granka, Pan, Hembrooke, and
  Gay}{Joachims et~al\mbox{.}}{2005}]%
        {joachims2005accurately}
\bibfield{author}{\bibinfo{person}{Thorsten Joachims}, \bibinfo{person}{Laura~A
  Granka}, \bibinfo{person}{Bing Pan}, \bibinfo{person}{Helene Hembrooke},
  {and} \bibinfo{person}{Geri Gay}.} \bibinfo{year}{2005}\natexlab{}.
\newblock \showarticletitle{Accurately interpreting clickthrough data as
  implicit feedback}. In \bibinfo{booktitle}{\emph{SIGIR}}.
\newblock


\bibitem[\protect\citeauthoryear{Joachims, Swaminathan, and Schnabel}{Joachims
  et~al\mbox{.}}{2017}]%
        {joachims2017unbiased}
\bibfield{author}{\bibinfo{person}{Thorsten Joachims}, \bibinfo{person}{Adith
  Swaminathan}, {and} \bibinfo{person}{Tobias Schnabel}.}
  \bibinfo{year}{2017}\natexlab{}.
\newblock \showarticletitle{Unbiased learning-to-rank with biased feedback}. In
  \bibinfo{booktitle}{\emph{WSDM}}.
\newblock


\bibitem[\protect\citeauthoryear{Li, Ren, Chen, Ren, Lian, and Ma}{Li
  et~al\mbox{.}}{2017}]%
        {li2017neural}
\bibfield{author}{\bibinfo{person}{Jing Li}, \bibinfo{person}{Pengjie Ren},
  \bibinfo{person}{Zhumin Chen}, \bibinfo{person}{Zhaochun Ren},
  \bibinfo{person}{Tao Lian}, {and} \bibinfo{person}{Jun Ma}.}
  \bibinfo{year}{2017}\natexlab{}.
\newblock \showarticletitle{Neural attentive session-based recommendation}. In
  \bibinfo{booktitle}{\emph{CIKM}}.
\newblock


\bibitem[\protect\citeauthoryear{Liu, Zeng, Mokhosi, and Zhang}{Liu
  et~al\mbox{.}}{2018}]%
        {liu2018stamp}
\bibfield{author}{\bibinfo{person}{Qiao Liu}, \bibinfo{person}{Yifu Zeng},
  \bibinfo{person}{Refuoe Mokhosi}, {and} \bibinfo{person}{Haibin Zhang}.}
  \bibinfo{year}{2018}\natexlab{}.
\newblock \showarticletitle{STAMP: short-term attention/memory priority model
  for session-based recommendation}. In \bibinfo{booktitle}{\emph{KDD}}.
\newblock


\bibitem[\protect\citeauthoryear{Ma, Zhao, Yi, Chen, Hong, and Chi}{Ma
  et~al\mbox{.}}{2018b}]%
        {ma2018modeling}
\bibfield{author}{\bibinfo{person}{Jiaqi Ma}, \bibinfo{person}{Zhe Zhao},
  \bibinfo{person}{Xinyang Yi}, \bibinfo{person}{Jilin Chen},
  \bibinfo{person}{Lichan Hong}, {and} \bibinfo{person}{Ed~H Chi}.}
  \bibinfo{year}{2018}\natexlab{b}.
\newblock \showarticletitle{Modeling task relationships in multi-task learning
  with multi-gate mixture-of-experts}. In \bibinfo{booktitle}{\emph{KDD}}.
\newblock


\bibitem[\protect\citeauthoryear{Ma, Zhao, Huang, Wang, Hu, Zhu, and Gai}{Ma
  et~al\mbox{.}}{2018a}]%
        {ma2018entire}
\bibfield{author}{\bibinfo{person}{Xiao Ma}, \bibinfo{person}{Liqin Zhao},
  \bibinfo{person}{Guan Huang}, \bibinfo{person}{Zhi Wang},
  \bibinfo{person}{Zelin Hu}, \bibinfo{person}{Xiaoqiang Zhu}, {and}
  \bibinfo{person}{Kun Gai}.} \bibinfo{year}{2018}\natexlab{a}.
\newblock \showarticletitle{Entire space multi-task model: An effective
  approach for estimating post-click conversion rate}. In
  \bibinfo{booktitle}{\emph{SIGIR}}.
\newblock


\bibitem[\protect\citeauthoryear{Mei and Eisner}{Mei and Eisner}{2016}]%
        {mei2016neural}
\bibfield{author}{\bibinfo{person}{Hongyuan Mei} {and} \bibinfo{person}{Jason
  Eisner}.} \bibinfo{year}{2016}\natexlab{}.
\newblock \showarticletitle{The neural hawkes process: A neurally
  self-modulating multivariate point process}.
\newblock \bibinfo{journal}{\emph{arXiv preprint arXiv:1612.09328}}
  (\bibinfo{year}{2016}).
\newblock


\bibitem[\protect\citeauthoryear{Meng, Yang, and Xiao}{Meng
  et~al\mbox{.}}{2020}]%
        {meng2020incorporating}
\bibfield{author}{\bibinfo{person}{Wenjing Meng}, \bibinfo{person}{Deqing
  Yang}, {and} \bibinfo{person}{Yanghua Xiao}.}
  \bibinfo{year}{2020}\natexlab{}.
\newblock \showarticletitle{Incorporating user micro-behaviors and item
  knowledge into multi-task learning for session-based recommendation}. In
  \bibinfo{booktitle}{\emph{SIGIR}}.
\newblock


\bibitem[\protect\citeauthoryear{Ni, Ou, Liu, Li, Ou, Zeng, and Si}{Ni
  et~al\mbox{.}}{2018}]%
        {ni2018perceive}
\bibfield{author}{\bibinfo{person}{Yabo Ni}, \bibinfo{person}{Dan Ou},
  \bibinfo{person}{Shichen Liu}, \bibinfo{person}{Xiang Li},
  \bibinfo{person}{Wenwu Ou}, \bibinfo{person}{Anxiang Zeng}, {and}
  \bibinfo{person}{Luo Si}.} \bibinfo{year}{2018}\natexlab{}.
\newblock \showarticletitle{Perceive your users in depth: Learning universal
  user representations from multiple e-commerce tasks}. In
  \bibinfo{booktitle}{\emph{KDD}}.
\newblock


\bibitem[\protect\citeauthoryear{Ren, Fang, Zhang, Liu, Li, Zhang, Yu, and
  Wang}{Ren et~al\mbox{.}}{2018}]%
        {ren2018learning}
\bibfield{author}{\bibinfo{person}{Kan Ren}, \bibinfo{person}{Yuchen Fang},
  \bibinfo{person}{Weinan Zhang}, \bibinfo{person}{Shuhao Liu},
  \bibinfo{person}{Jiajun Li}, \bibinfo{person}{Ya Zhang},
  \bibinfo{person}{Yong Yu}, {and} \bibinfo{person}{Jun Wang}.}
  \bibinfo{year}{2018}\natexlab{}.
\newblock \showarticletitle{Learning multi-touch conversion attribution with
  dual-attention mechanisms for online advertising}. In
  \bibinfo{booktitle}{\emph{CIKM}}.
\newblock


\bibitem[\protect\citeauthoryear{Ren, Qin, Zheng, Yang, Zhang, Qiu, and Yu}{Ren
  et~al\mbox{.}}{2019}]%
        {ren2019deep}
\bibfield{author}{\bibinfo{person}{Kan Ren}, \bibinfo{person}{Jiarui Qin},
  \bibinfo{person}{Lei Zheng}, \bibinfo{person}{Zhengyu Yang},
  \bibinfo{person}{Weinan Zhang}, \bibinfo{person}{Lin Qiu}, {and}
  \bibinfo{person}{Yong Yu}.} \bibinfo{year}{2019}\natexlab{}.
\newblock \showarticletitle{Deep Recurrent Survival Analysis}. In
  \bibinfo{booktitle}{\emph{AAAI}}.
\newblock


\bibitem[\protect\citeauthoryear{Richardson, Dominowska, and Ragno}{Richardson
  et~al\mbox{.}}{2007}]%
        {richardson2007predicting}
\bibfield{author}{\bibinfo{person}{Matthew Richardson}, \bibinfo{person}{Ewa
  Dominowska}, {and} \bibinfo{person}{Robert Ragno}.}
  \bibinfo{year}{2007}\natexlab{}.
\newblock \showarticletitle{Predicting clicks: estimating the click-through
  rate for new ads}. In \bibinfo{booktitle}{\emph{WWW}}.
\newblock


\bibitem[\protect\citeauthoryear{Tang, Liu, Zhao, and Gong}{Tang
  et~al\mbox{.}}{2020}]%
        {tang2020progressive}
\bibfield{author}{\bibinfo{person}{Hongyan Tang}, \bibinfo{person}{Junning
  Liu}, \bibinfo{person}{Ming Zhao}, {and} \bibinfo{person}{Xudong Gong}.}
  \bibinfo{year}{2020}\natexlab{}.
\newblock \showarticletitle{Progressive layered extraction (ple): A novel
  multi-task learning (mtl) model for personalized recommendations}. In
  \bibinfo{booktitle}{\emph{Fourteenth ACM Conference on Recommender Systems}}.
\newblock


\bibitem[\protect\citeauthoryear{Wang, Chang, Liu, Huang, Chen, Yu, Li, and
  Chu}{Wang et~al\mbox{.}}{2022}]%
        {wang2022escm}
\bibfield{author}{\bibinfo{person}{Hao Wang}, \bibinfo{person}{Tai-Wei Chang},
  \bibinfo{person}{Tianqiao Liu}, \bibinfo{person}{Jianmin Huang},
  \bibinfo{person}{Zhichao Chen}, \bibinfo{person}{Chao Yu},
  \bibinfo{person}{Ruopeng Li}, {and} \bibinfo{person}{Wei Chu}.}
  \bibinfo{year}{2022}\natexlab{}.
\newblock \showarticletitle{ESCM2: Entire Space Counterfactual Multi-Task Model
  for Post-Click Conversion Rate Estimation}.
\newblock \bibinfo{journal}{\emph{SIGIR}} (\bibinfo{year}{2022}).
\newblock


\bibitem[\protect\citeauthoryear{Wang, Lin, Lin, Yang, and Wu}{Wang
  et~al\mbox{.}}{2020}]%
        {wang2020m2grl}
\bibfield{author}{\bibinfo{person}{Menghan Wang}, \bibinfo{person}{Yujie Lin},
  \bibinfo{person}{Guli Lin}, \bibinfo{person}{Keping Yang}, {and}
  \bibinfo{person}{Xiao-ming Wu}.} \bibinfo{year}{2020}\natexlab{}.
\newblock \showarticletitle{M2GRL: A Multi-task Multi-view Graph Representation
  Learning Framework for Web-scale Recommender Systems}. In
  \bibinfo{booktitle}{\emph{KDD}}.
\newblock


\bibitem[\protect\citeauthoryear{Wang, Bendersky, Metzler, and Najork}{Wang
  et~al\mbox{.}}{2016}]%
        {wang2016learning}
\bibfield{author}{\bibinfo{person}{Xuanhui Wang}, \bibinfo{person}{Michael
  Bendersky}, \bibinfo{person}{Donald Metzler}, {and} \bibinfo{person}{Marc
  Najork}.} \bibinfo{year}{2016}\natexlab{}.
\newblock \showarticletitle{Learning to rank with selection bias in personal
  search}. In \bibinfo{booktitle}{\emph{SIGIR}}.
\newblock


\bibitem[\protect\citeauthoryear{Wang, Golbandi, Bendersky, Metzler, and
  Najork}{Wang et~al\mbox{.}}{2018}]%
        {wang2018position}
\bibfield{author}{\bibinfo{person}{Xuanhui Wang}, \bibinfo{person}{Nadav
  Golbandi}, \bibinfo{person}{Michael Bendersky}, \bibinfo{person}{Donald
  Metzler}, {and} \bibinfo{person}{Marc Najork}.}
  \bibinfo{year}{2018}\natexlab{}.
\newblock \showarticletitle{Position bias estimation for unbiased learning to
  rank in personal search}. In \bibinfo{booktitle}{\emph{WSDM}}.
\newblock


\bibitem[\protect\citeauthoryear{Wen, Zhang, Lv, Bao, Wang, and Chen}{Wen
  et~al\mbox{.}}{2021}]%
        {wen2021hierarchically}
\bibfield{author}{\bibinfo{person}{Hong Wen}, \bibinfo{person}{Jing Zhang},
  \bibinfo{person}{Fuyu Lv}, \bibinfo{person}{Wentian Bao},
  \bibinfo{person}{Tianyi Wang}, {and} \bibinfo{person}{Zulong Chen}.}
  \bibinfo{year}{2021}\natexlab{}.
\newblock \showarticletitle{Hierarchically Modeling Micro and Macro Behaviors
  via Multi-Task Learning for Conversion Rate Prediction}.
\newblock \bibinfo{journal}{\emph{SIGIR}} (\bibinfo{year}{2021}).
\newblock


\bibitem[\protect\citeauthoryear{Wen, Zhang, Wang, Lv, Bao, Lin, and Yang}{Wen
  et~al\mbox{.}}{2020}]%
        {wen2019entire}
\bibfield{author}{\bibinfo{person}{Hong Wen}, \bibinfo{person}{Jing Zhang},
  \bibinfo{person}{Yuan Wang}, \bibinfo{person}{Fuyu Lv},
  \bibinfo{person}{Wentian Bao}, \bibinfo{person}{Quan Lin}, {and}
  \bibinfo{person}{Keping Yang}.} \bibinfo{year}{2020}\natexlab{}.
\newblock \showarticletitle{Entire Space Multi-Task Modeling via Post-Click
  Behavior Decomposition for Conversion Rate Prediction}.
\newblock \bibinfo{journal}{\emph{SIGIR}}.
\newblock


\bibitem[\protect\citeauthoryear{Wu, Ahmed, Beutel, Smola, and Jing}{Wu
  et~al\mbox{.}}{2017}]%
        {wu2017recurrent}
\bibfield{author}{\bibinfo{person}{Chao-Yuan Wu}, \bibinfo{person}{Amr Ahmed},
  \bibinfo{person}{Alex Beutel}, \bibinfo{person}{Alexander~J Smola}, {and}
  \bibinfo{person}{How Jing}.} \bibinfo{year}{2017}\natexlab{}.
\newblock \showarticletitle{Recurrent recommender networks}. In
  \bibinfo{booktitle}{\emph{WSDM}}.
\newblock


\bibitem[\protect\citeauthoryear{Xi, Chen, Yan, Zhang, Zhu, Zhuang, and
  Chen}{Xi et~al\mbox{.}}{2021}]%
        {xi2021modeling}
\bibfield{author}{\bibinfo{person}{Dongbo Xi}, \bibinfo{person}{Zhen Chen},
  \bibinfo{person}{Peng Yan}, \bibinfo{person}{Yinger Zhang},
  \bibinfo{person}{Yongchun Zhu}, \bibinfo{person}{Fuzhen Zhuang}, {and}
  \bibinfo{person}{Yu Chen}.} \bibinfo{year}{2021}\natexlab{}.
\newblock \showarticletitle{Modeling the Sequential Dependence among Audience
  Multi-step Conversions with Multi-task Learning in Targeted Display
  Advertising}.
\newblock \bibinfo{journal}{\emph{KDD}} (\bibinfo{year}{2021}).
\newblock


\bibitem[\protect\citeauthoryear{Zhang, Bao, Liu, Yang, Lin, Wen, and
  Ramezani}{Zhang et~al\mbox{.}}{2020}]%
        {zhang2020large}
\bibfield{author}{\bibinfo{person}{Wenhao Zhang}, \bibinfo{person}{Wentian
  Bao}, \bibinfo{person}{Xiao-Yang Liu}, \bibinfo{person}{Keping Yang},
  \bibinfo{person}{Quan Lin}, \bibinfo{person}{Hong Wen}, {and}
  \bibinfo{person}{Ramin Ramezani}.} \bibinfo{year}{2020}\natexlab{}.
\newblock \showarticletitle{Large-scale causal approaches to debiasing
  post-click conversion rate estimation with multi-task learning}. In
  \bibinfo{booktitle}{\emph{WWW}}.
\newblock


\bibitem[\protect\citeauthoryear{Zhao, Cheng, Hong, and Chi}{Zhao
  et~al\mbox{.}}{2015}]%
        {zhao2015improving}
\bibfield{author}{\bibinfo{person}{Zhe Zhao}, \bibinfo{person}{Zhiyuan Cheng},
  \bibinfo{person}{Lichan Hong}, {and} \bibinfo{person}{Ed~H Chi}.}
  \bibinfo{year}{2015}\natexlab{}.
\newblock \showarticletitle{Improving user topic interest profiles by behavior
  factorization}. In \bibinfo{booktitle}{\emph{WWW}}.
\newblock


\bibitem[\protect\citeauthoryear{Zhao, Hong, Wei, Chen, Nath, Andrews,
  Kumthekar, Sathiamoorthy, Yi, and Chi}{Zhao et~al\mbox{.}}{2019}]%
        {zhao2019recommending}
\bibfield{author}{\bibinfo{person}{Zhe Zhao}, \bibinfo{person}{Lichan Hong},
  \bibinfo{person}{Li Wei}, \bibinfo{person}{Jilin Chen},
  \bibinfo{person}{Aniruddh Nath}, \bibinfo{person}{Shawn Andrews},
  \bibinfo{person}{Aditee Kumthekar}, \bibinfo{person}{Maheswaran
  Sathiamoorthy}, \bibinfo{person}{Xinyang Yi}, {and} \bibinfo{person}{Ed
  Chi}.} \bibinfo{year}{2019}\natexlab{}.
\newblock \showarticletitle{Recommending what video to watch next: a multitask
  ranking system}. In \bibinfo{booktitle}{\emph{RecSys}}.
\newblock


\end{thebibliography}

\end{document}